\RequirePackage{fix-cm}
\documentclass[twocolumn,epjc3]{svjour3}
\smartqed  
\RequirePackage{graphicx}
\usepackage{hyperref}
\usepackage{fouridx}
\usepackage{epstopdf}
\usepackage{amsmath}
\usepackage{hyperref}
\journalname{Eur. Phys. J. C}
\begin{document}

\title{Angular distributions of the polarized photons and electron in the decays of the $^3D_3$ state of charmonium}

\author{Alex W. K. Mok\thanksref{e1,addr1}
        \and
        Cheuk-Ping Wong\thanksref{e2,addr1}
        \and
        Wai-Yu Sit\thanksref{e3,addr1}
}

\thankstext{e1}{e-mail: wkmok@hkbu.edu.hk}
\thankstext{e2}{e-mail: cheuk-ping.wong@polyu.edu.hk}
\thankstext{e3}{e-mail: 12466654@mail.life.hkbu.edu.hk}

\institute{Department of Physics, Hong Kong Baptist University, Kowloon Tong, Hong Kong, China\label{addr1}}

\date{Received: date / Accepted: date}

\maketitle

\begin{abstract}
We calculate the combined angular distribution functions of the polarized photons ($\gamma_1$ and $\gamma_2$) and electron ($e^-$) produced in the cascade process $\bar{p}p\rightarrow\fourIdx{3}{}{}{3}{D}\rightarrow\fourIdx{3}{}{}{2}{P}+\gamma_1
\rightarrow(\psi+\gamma_2)+\gamma_1\rightarrow(e^++e^-)+\gamma_1+\gamma_2$, when the colliding $\bar{p}$ and $p$ are unpolarized. Our results are independent of any dynamical models and are expressed in terms of the spherical harmonics whose coefficients are functions of the angular-momentum helicity amplitudes of the individual processes. Once the joint angular distribution of ($\gamma_1$, $\gamma_2$) and that of ($\gamma_2$, $e^-$) with the polarization of either one of the two particles are measured, our results will enable one to determine the relative magnitudes as well as the relative phases of all the angular-momentum helicity amplitudes in the radiative decay processes $\fourIdx{3}{}{}{3}{D}\rightarrow\fourIdx{3}{}{}{2}{P}+\gamma_1$ and $\fourIdx{3}{}{}{2}{P}\rightarrow\psi+\gamma_2$.
\end{abstract}
\allowdisplaybreaks
\section{Introduction}
\label{intro}
Recently there has been great interest in charmonium spectroscopy above the open charm $D\bar{D}$ threshold of $3.73$ GeV \cite{Higher charmonia,New states above charm threshold,Quarkonia and their transitions,Charmonium spectroscopy above thresholds}. Although the mass of the unobserved $1\fourIdx{3}{}{}{3}{D}$ state of charmonium is expected to lie slightly above the charm threshold \cite{New states above charm threshold}, its Zweig-allowed strong decay to $D\bar{D}$ is suppressed by the $F$-wave angular momentum barrier \cite{Charmonium options for the X(3872),Charmonium levels near threshold and the narrow state}. The total strong width of $1\fourIdx{3}{}{}{3}{D}$ is predicted to be just $0.5$ MeV \cite{Higher charmonia} and therefore other decay modes such as $\gamma\fourIdx{3}{}{}{2}{P}$ and $\pi\pi J/\psi$ may be observable \cite{Quarkonia and their transitions}. The measurement of the angular distributions in these prominent radiative and hadronic decays of the charmonium $1\fourIdx{3}{}{}{3}{D}$ state can provide valuable information on the true dynamics of the charmonium system above the charm threshold. In fact, the observation of the radiative decays of the charmonium states below and above the charm threshold is an important component of the planned PANDA experiments at FAIR \cite{The PANDA experiment at FAIR,Physics performance report for PANDA: strong interaction studies with antiprotons}, which study charmonium spectroscopy in $\bar{p}p$ annihilation.

In our previous paper \cite{previous paper}, it is shown that by measuring the combined angular distribution of the two photons ($\gamma_1$, $\gamma_2$) and that of the second photon and electron ($\gamma_2$, $e^-$), regardless of their polarizations, in the sequential decay process originating from unpolarized $\bar{p}p$ collisions, namely, $\bar{p}p\rightarrow\fourIdx{3}{}{}{3}{D}
\rightarrow\fourIdx{3}{}{}{2}{P}+\gamma_1\rightarrow(\psi+\gamma_2)+\gamma_1
\rightarrow(e^++e^-)+\gamma_1+\gamma_2$, one can extract the relative magnitudes as well as the cosines of the relative phases of all the angular-momentum helicity amplitudes in the radiative decay processes $\fourIdx{3}{}{}{3}{D}\rightarrow\fourIdx{3}{}{}{2}{P}+\gamma_1$ and $\fourIdx{3}{}{}{2}{P}\rightarrow\psi+\gamma_2$. The sines of the relative phases of these helicity amplitudes, however, cannot be determined uniquely. By including the measurement of the polarization of one of the decay particles, one may also obtain unambiguously the sines of the relative phases \cite{Mok 2010,Mok 2011}. So in this paper we calculate the combined angular distributions of the final particles ($\gamma_1$, $\gamma_2$ and $e^-$) with the determination of the polarization of one particle in the above cascade process when $\bar{p}$ and $p$ are unpolarized.

In general, the helicity amplitudes are complex and their relative phases are nontrivial. It is important to obtain them from experiments because we can then learn about the true dynamics of the charmonium system from the decays of the charmonium states. Once the combined angular distribution of $\gamma_1$, $\gamma_2$ and $e^-$ and the polarization of any one of the particles in unpolarized $\bar{p}p$ collisions are experimentally measured, our expressions will enable one to calculate the relative magnitudes as well as the relative phases of all the angular-momentum helicity amplitudes in the two radiative decay processes $\fourIdx{3}{}{}{3}{D}\rightarrow\fourIdx{3}{}{}{2}{P}+\gamma_1$ and $\fourIdx{3}{}{}{2}{P}\rightarrow\psi+\gamma_2$. As our calculation is based only on the general principles of quantum mechanics and symmetry, our results are independent of any dynamical models. In addition, our results on the partially integrated angular distributions where the combined angular distribution function of $\gamma_1$, $\gamma_2$ and $e^-$ is integrated over the direction of one of the three particles are quite interesting. They show that by measuring the two-particle angular distribution of ($\gamma_1$, $\gamma_2$) and that of ($\gamma_2$, $e^-$) with the polarization of either one of the two particles, one can also get complete information on the helicity amplitudes.

The format of the rest of this paper is as follows. In Sect.~\ref{sec:1}, we give the calculations for the combined angular distribution with polarization determination of the electron and of the two photons in the cascade process $\bar{p}p\rightarrow\fourIdx{3}{}{}{3}{D}\rightarrow\fourIdx{3}{}{}{2}{P}+\gamma_1
\rightarrow(\psi+\gamma_2)+\gamma_1\rightarrow(e^++e^-)+\gamma_1+\gamma_2$, when $\bar{p}$ and $p$ are unpolarized. We then show how the measurement of this joint angular distribution of polarized $\gamma_1$, $\gamma_2$ and $e^-$ enables us to obtain complete information on the helicity amplitudes in the two radiative decay processes $\fourIdx{3}{}{}{3}{D}\rightarrow\fourIdx{3}{}{}{2}{P}+\gamma_1$ and $\fourIdx{3}{}{}{2}{P}\rightarrow\psi+\gamma_2$. We also present three different results for the combined angular distribution, in which the polarization of only one of the three particles, $\gamma_1$, $\gamma_2$ and $e^-$, is measured. In Sect.~\ref{sec:2}, we present the results for the partially integrated angular distributions in different cases where the combined angular distribution function of the three particles is integrated over the direction of one particle. These results can all be expressed in terms of the orthogonal spherical harmonic functions. We point out how the measurement of these two-particle angular distributions will again give us complete information on all the helicity amplitudes in the two radiative decay processes. Finally, in Sect.~\ref{sec:3}, we make some concluding remarks.
\section{The polarized angular distributions of the photons and electron}
\label{sec:1}
\begin{figure}
 \includegraphics[scale=0.4]{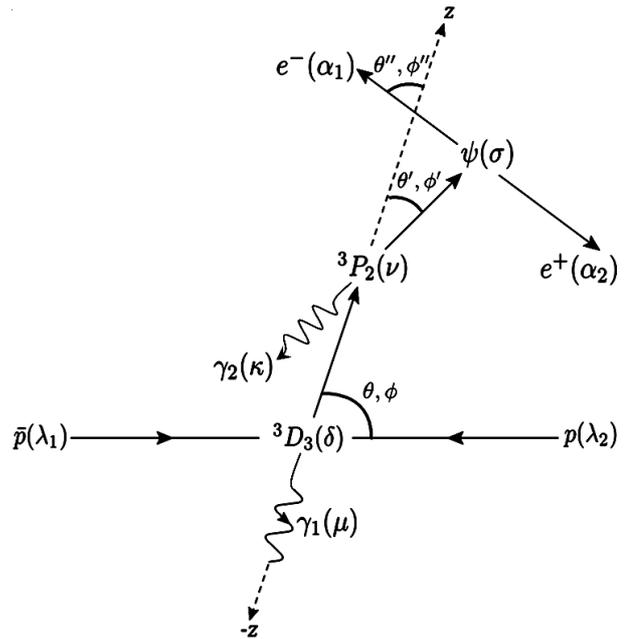}
\caption{Symbolic sketch of   $\bar{p}(\lambda_1)+p(\lambda_2)\rightarrow\fourIdx{3}{}{}{3}{D}(\delta)\rightarrow\fourIdx{3}{}{}{2}{P}(\nu)+\gamma_1(\mu)
  \rightarrow[\psi(\sigma)+\gamma_2(\kappa)]+\gamma_1(\mu)\rightarrow[e^-(\alpha_1)+e^+(\alpha_2)]
  +\gamma_1(\mu)+\gamma_2(\kappa)$ showing different angles of the decay particles.}
\label{graph}       
\end{figure}

We consider the cascade process, $\bar{p}(\lambda_1)+p(\lambda_2)\rightarrow
\fourIdx{3}{}{}{3}{D}(\delta)$ $\rightarrow\fourIdx{3}{}{}{2}{P}(\nu)+\gamma_1(\mu)
\rightarrow[\psi(\sigma)+\gamma_2(\kappa)]+\gamma_1(\mu)
\rightarrow[e^-(\alpha_1)+e^+(\alpha_2)]+\gamma_1(\mu)+\gamma_2(\kappa)$, in the $\fourIdx{3}{}{}{3}{D}$ rest frame or the $\bar{p}p$ c.m. frame. The Greek symbols in the brackets represent the helicities of the particles except $\delta$ which represents the $z$ component of the angular momentum of the stationary $\fourIdx{3}{}{}{3}{D}$ resonance. We choose the $z$ axis to be the direction of motion of $\fourIdx{3}{}{}{2}{P}$ in the $\fourIdx{3}{}{}{3}{D}$ rest frame. The $x$ and $y$ axes are arbitrary and the experimentalists can choose them according to their convenience. A symbolic sketch of the cascade process is shown in Fig.~\ref{graph}.

Following the conventions of our previous paper \cite{previous paper}, the probability amplitude for the cascade process can be expressed in terms of the Wigner $D$-functions and the angular-momentum helicity amplitudes $B_{\lambda_1\lambda_2}$, $A_{\nu\mu}$, $E_{\alpha\kappa}$ and $C_{\alpha_1\alpha_2}$ for the individual sequential processes as
\begin{eqnarray}\label{sequential process T}
T^{\alpha_1\alpha_2\mu\kappa}_{\lambda_1\lambda_2}
&=&\frac{7\sqrt{15}}{16\pi^2}C_{\alpha_1\alpha_2}B_{\lambda_1\lambda_2}
    \sum^{-3\rightarrow3}_{\delta}\sum^{-1\rightarrow1}_{\sigma}A_{\mu+\delta,\mu}E_{\sigma\kappa}\nonumber\\
    &\quad&\times D^3_{\delta\lambda}(\phi,\theta,-\phi)
        D^{2*}_{\mu+\delta,\sigma-\kappa}(\phi',\theta',-\phi')\nonumber\\
    &\quad&\times D^{1*}_{\sigma\alpha}(\phi'',\theta'',-\phi'').
\end{eqnarray}
In the $D$-functions, the angles $(\phi,\theta)$ giving the direction of $\bar{p}$, the angles $(\phi',\theta')$ giving the direction of $\psi$ and the angles $(\phi'',\theta'')$ giving the direction of $e^-$ are measured in the $\fourIdx{3}{}{}{3}{D}$, the $\fourIdx{3}{}{}{2}{P}$ and the $\psi$ rest frames, respectively. The angles of each decay particle observed in different rest frames can be calculated using the Lorentz transformation. The equations relating these angles are given in \cite{Mok 2008}.

Because of the C and P invariances, the angular-momentum helicity amplitudes in (\ref{sequential process T}) are not all independent. We have
\begin{eqnarray}\label{after using C and P}
B_{\lambda_1\lambda_2}&\overset{\text{P}}{=}&B_{-\lambda_1-\lambda_2},\nonumber\\
B_{\lambda_1\lambda_2}&\overset{\text{C}}{=}&B_{\lambda_2\lambda_1},\nonumber\\
A_{\nu\mu}&\overset{\text{P}}{=}&A_{-\nu-\mu},\nonumber\\
E_{\sigma\kappa}&\overset{\text{P}}{=}&E_{-\sigma-\kappa},\nonumber\\
C_{\alpha_1\alpha_2}&\overset{\text{P}}{=}&C_{-\alpha_1-\alpha_2},\nonumber\\
C_{\alpha_1\alpha_2}&\overset{\text{C}}{=}&C_{\alpha_2\alpha_1}.
\end{eqnarray}
Making use of the symmetry relations of (\ref{after using C and P}), we now re-label the independent angular-momentum helicity amplitudes as follows:
\begin{eqnarray}
B_0&=&\sqrt{2}B_{\frac{1}{2}\frac{1}{2}},\qquad
    B_1=\sqrt{2}B_{\frac{1}{2}-\frac{1}{2}},\nonumber\\
A_i&=&A_{i-2,1}=A_{2-i,-1}\qquad(i=0,1,2,3,4),\nonumber\\
E_j&=&E_{j-1,1}=E_{1-j,-1}\qquad(j=0,1,2),\nonumber\\
C_0&=&\sqrt{2}C_{\frac{1}{2}\frac{1}{2}},\qquad C_1=\sqrt{2}C_{\frac{1}{2}-\frac{1}{2}}.
\end{eqnarray}
We will also make use of the following normalizations:
\begin{eqnarray}
|B_0|^2+|B_1|^2=|C_0|^2+|C_1|^2=1
\end{eqnarray}
\noindent
and
\begin{eqnarray}
\sum^4_{i=0}|A_i|^2=\sum^2_{j=0}|E_j|^2=1
\end{eqnarray}

When $\bar{p}$ and $p$ are unpolarized, the normalized function describing the combined angular distribution of the electron and the two photons whose polarizations are also observed can be written as
\begin{eqnarray}\label{normalized function}
&W_{\mu\kappa\alpha_1}&(\theta,\phi;\theta',\phi';\theta'',\phi'')\nonumber\\
&\quad&=N\sum^{\pm\frac{1}{2}}_{\lambda_1,\lambda_2}\sum^{\pm\frac{1}{2}}_{\alpha_2}
T^{\alpha_1\alpha_2\mu\kappa}_{\lambda_1\lambda_2}
T^{\alpha_1\alpha_2\mu\kappa*}_{\lambda_1\lambda_2},
\end{eqnarray}
where the subscripts $\mu$, $\kappa$ and $\alpha_1$ of $W$ represent the polarizations of $\gamma_1$, $\gamma_2$ and $e^-$, respectively. The normalization constant $N$ in (\ref{normalized function}) is determined by requiring that the integral of the angular distribution function $W_{\mu\kappa\alpha_1}(\theta,\phi;\theta',\phi';\theta'',\phi'')$ over all the directions of $\gamma_1$, $\gamma_2$ and $e^-$ or over all the angles, $(\theta,\phi;\theta',\phi';\theta'',\phi'')$, is $1$. In (\ref{normalized function}) we sum over the helicities $\alpha_2$ since $e^+$ is not observed. Substituting (\ref{sequential process T}) into (\ref{normalized function}) and performing the various sums will then give us an expression for the angular distribution function $W_{\mu\kappa\alpha_1}(\theta,\phi;\theta',\phi';\theta'',\phi'')$ in terms of the Wigner $D$-functions. After very long algebra, we get
\begin{align}\label{wigner final_pol.}
&W_{\mu\kappa\alpha_1}(\theta,\phi;\theta',\phi';\theta'',\phi'')\nonumber\\
    &\quad=\frac{1}{(4\pi)^3}\sum^{0,2,4,6}_{J_1}B^{J_1}\sum^{2}_{J_3=0}C^{J_3}_{\alpha_1}\nonumber\\
    &\qquad\times\sum^{4}_{J_2=0}(-1)^{\left(\frac{2+\mu-\kappa}{2}\right)J_2+\left(\frac{1-\kappa}{2}\right)J_3}\nonumber\\
        &\qquad\times\sum^{d_m}_{d=0}\sum^{d_m'}_{d'=0}
        \left(1-\frac{\delta_{d0}}{2}\right)\left(1-\frac{\delta_{d'0}}{2}\right)I^{J_1J_2J_3}_{dd'}
\end{align}
where
\begin{align}
    B^{J_1}&=\sqrt{7}\sum^{1}_{\lambda=0}(-1)^\lambda\langle33;\lambda,-\lambda|J_1;0\rangle|B_\lambda|^2,\\
    C^{J_3}_{\alpha_1}&=(-1)^{\left(\alpha_1+\frac{1}{2}\right)J_3}\sqrt{3}\nonumber\\
        &\quad\times\sum^{1}_{\alpha=0}(-1)^\alpha
        \langle11;\alpha,-\alpha|J_3;0\rangle|C_\alpha|^2,\\
    I^{J_1J_2J_3}_{dd'}&=\beta^{J_1J_2}_{d+}\left[\gamma^{J_3J_2}_{d'+}(D_1+D^*_1+D_2+D^*_2)\right.\nonumber\\
        &\quad\left.+\gamma^{J_3J_2}_{d'-}(D_1-D^*_1+D_2-D^*_2)\right]\nonumber\\
        &\quad+\beta^{J_1J_2}_{d-}\left[\gamma^{J_3J_2}_{d'+}(D_1-D^*_1-D_2+D^*_2)\right.\nonumber\\
        &\quad\left.+\gamma^{J_3J_2}_{d'-}(D_1+D^*_1-D_2-D^*_2)\right],\\
    \label{definition beta}
    \beta^{J_1J_2}_{d\pm}&=\frac{\sqrt{35}}{2}\sum_{s(d)}
        \left(A_{\frac{s+d}{2}}A^*_{\frac{s-d}{2}}\pm A^*_{\frac{s+d}{2}}A_{\frac{s-d}{2}}\right)\nonumber\\
        &\quad\times
        \left\langle22;\frac{s+d-4}{2},-\frac{s-d-4}{2}|J_2;d\right\rangle\nonumber\\
        &\quad\times\left\langle33;\frac{s+d-6}{2},-\frac{s-d-6}{2}|J_1;d\right\rangle,\\
    \label{definition gamma}
    \gamma^{J_3J_2}_{d'\pm}&=\frac{\sqrt{15}}{2}\sum_{s'(d')}
        \left(E_{\frac{s'+d'}{2}}E^*_{\frac{s'-d'}{2}}\pm E^*_{\frac{s'+d'}{2}}E_{\frac{s'-d'}{2}}\right)\nonumber\\
        &\quad\times
        \left\langle22;\frac{s'+d'}{2},-\frac{s'-d'}{2}|J_2;d'\right\rangle\nonumber\\
        &\quad\times\left\langle11;\frac{s'+d'-2}{2},-\frac{s'-d'-2}{2}|J_3;d'\right\rangle,\\
    D_1&=D^{J_1}_{\mu d,0}(\theta,\phi,-\theta)
        D^{J_2*}_{\mu d,-\kappa d'}(\theta',\phi',-\theta')\nonumber\\
        &\quad\times D^{J_3*}_{-\kappa d',0}(\theta'',\phi'',-\theta''),\label{D_1}\\
    D_2&=D^{J_1*}_{\mu d,0}(\theta,\phi,-\theta)
        D^{J_2}_{\mu d,\kappa d'}(\theta',\phi',-\theta')\nonumber\\
        &\quad\times D^{J_3}_{\kappa d',0}(\theta'',\phi'',-\theta''),\label{D_2}\\
    s(d)&=|d|,|d|+2,...,8-|d|,\\s'(d')&=|d'|,|d'|+2,...,4-|d'|,\\
    d_m&=\min\{4,J_1,J_2\},\\d_m'&=\min\{2,J_2,J_3\}.
    \end{align}The Wigner $D$-functions in (\ref{D_1}) and (\ref{D_2}) are given by \cite{Martin}
\begin{align}\label{Martin}
D^j_{m,m'}(\alpha,\beta,\gamma)=\langle j,m|R(\alpha,\beta,\gamma)|j,m'\rangle
\end{align}where $\alpha$, $\beta$, $\gamma$ are Euler angles and the rotation operator $R(\alpha,\beta,\gamma)$ can be writtern as
\begin{align}
R(\alpha,\beta,\gamma)=e^{-\operatorname{i}\alpha J_z}e^{-\operatorname{i}\beta J_y}e^{-\operatorname{i}\gamma J_z}.
\end{align}The explicit expressions for all the coefficients in (\ref{wigner final_pol.}) are given in \ref{sec:apdix}. Making use of the orthogonal relation of the Wigner $D$-functions,
\begin{align}
&\int^{2\pi}_{0}d\alpha\int^{2\pi}_{0}d\gamma\int^{\pi}_{0}D^{j*}_{mm'}(\alpha,\beta,\gamma)
    D^{j'}_{\mu\mu'}\operatorname{sin}\beta d\beta\nonumber\\
    &\quad=\frac{8\pi^2}{2j+1}\delta_{m\mu}\delta_{m'\mu'}\delta_{jj'},
\end{align} we can obtain these coefficients as
\begin{align}\label{coefficient function}
&(-1)^{\frac{J_2}{2}(2+\mu-\kappa)+\frac{J_3}{2}(1-\kappa)}B^{J_1}C^{J_3}_{\alpha_1}
    \Big\{\beta^{J_1J_2}_{d+}\nonumber\\
    &\quad\times\left[\gamma^{J_3J_2}_{d'+}+\gamma^{J_3J_2}_{d'-}(1-\delta_{d'0})\right]\nonumber\\
    &\quad+\beta^{J_1J_2}_{d-}\left[\gamma^{J_3J_2}_{d'+}(1-\delta_{d0})
    +\gamma^{J_3J_2}_{d'-}(1-\delta_{d0})(1-\delta_{d'0})\right]\Big\}\nonumber\\
&=(2J_1+1)(2J_2+1)(2J_3+1)\nonumber\\
    &\quad\times\int
    W_{\mu\kappa\alpha_1}(\theta,\phi;\theta',\phi';\theta'',\phi'')D^*_1d\Omega d\Omega'd\Omega''.
\end{align}

When we have sufficient experimental data for the angular distribution function $W_{\mu\kappa\alpha}$, where the final polarizations, $\mu$, $\kappa$ and $\alpha$, of all the three decay particles are measured, the integral on the right side of (\ref{coefficient function}) can be determined numerically for all possible allowed values of $J_1$, $J_2$, $J_3$, $d$ and $d'$. Thus we can extract the different coefficients $B^{J_1}$, $C^{J_3}_{\alpha_1}$, $\beta^{J_1J_2}_{d\pm}$ and $\gamma^{J_3J_2}_{d'\pm}$ on the left side of (\ref{coefficient function}). From these coefficients we can determine the relative magnitudes of the $A$, $B$, $C$ and $E$ helicity amplitudes as well as the cosines and sines of the relative phases of the $A$ and $E$ helicity amplitudes in the radiative decay processes $\fourIdx{3}{}{}{3}{D}\rightarrow\fourIdx{3}{}{}{2}{P}+\gamma_1$ and $\fourIdx{3}{}{}{2}{P}\rightarrow\psi+\gamma_2$, respectively. Let us illustrate more clearly how the measurements of $(J_1J_2J_3dd')$ coefficients can give us all the information. First, the measurement of the $(00100)$ and $(00200)$ coefficients yields $\gamma^{10}_{0+}$ and $\gamma^{20}_{0+}$, and with the normalization $|E_0|^2+|E_1|^2+|E_2|^2=1$, the relative magnitudes of $E_j$ are determined. Next the measurements of $(01000)$, $(02000)$, $(03000)$ and $(04000)$ coefficients yields $\beta^{01}_{0+}$, $\beta^{02}_{0+}$, $\beta^{03}_{0+}$ and $\beta^{04}_{0+}$, and with the normalization $|A_0|^2+|A_1|^2+|A_2|^2+|A_3|^2+|A_4|^2=1$, the relative magnitudes of $A_i$ are determined. Measuring $(20000)$ yields $B^2$ and with the normalization $|B_0|^2+|B_1|^2=1$, the relative magnitudes of $B_0$ and $B_1$ are obtained. The measurement of $(00200)$ yields $C^2_{\pm\frac{1}{2}}$ and with the normalization $|C_0|^2+|C_1|^2$, the relative magnitudes of $C_0$ and $C_1$ are determined. After having obtained all the relative magnitudes, now measuring the $(02101)$ and $(02201)$ coefficients yields $\operatorname{Re}(E_1E^*_0)$, $\operatorname{Re}(E_2E^*_1)$, $\operatorname{Im}(E_1E^*_0)$ and $\operatorname{Im}(E_2E^*_1)$. Hence the cosines and sines of the relative phases of $E_j$ are determined. Finally, by measuring the $(22010)$, $(24010)$, $(42010)$ and $(44010)$ coefficients, we can obtain the cosines and sines of the relative phases of $A_i$.

By summing over one or two helicity indices ($\mu$, $\kappa$ and $\alpha_1$) of (\ref{wigner final_pol.}), we can easily obtain different combined angular distribution functions where the polarizations of only one or two decay products ($\gamma_1$, $\gamma_2$ and $e^-$) are measured. Suppose we are interested in only measuring the polarization $\mu$ of $\gamma_1$, the normalized combined angular distribution of $\gamma_1$, $\gamma_2$ and $e^-$ will then become
\begin{align}\label{W_mu}
&W_\mu(\theta,\phi;\theta',\phi';\theta'',\phi'')\nonumber\\
    &\quad=\frac{1}{4}\sum^{\pm1}_{\kappa}\sum^{\pm\frac{1}{2}}_{\alpha_1}
        W_{\mu\kappa\alpha_1}(\theta,\phi;\theta',\phi';\theta'',\phi'')\nonumber\\
    &\quad=\frac{1}{2(4\pi)^3}\sum^{0,2,4,6}_{J_1}B^{J_1}\sum^{0,2}_{J_3}C^{J_3}
        \sum^{4}_{J_2=0}(-1)^{\left(\frac{1+\mu}{2}\right)J_2}\nonumber\\
        &\qquad\times\sum^{d_m}_{d=0}\sum^{d_m'}_{d'=0}
        \left(1-\frac{\delta_{d0}}{2}\right)\left(1-\frac{\delta_{d'0}}{2}\right)
        \Bigg\{\beta^{J_1J_2}_{d+}\nonumber\\
    &\qquad\times\Big\{\gamma^{J_3J_2}_{d'+}\left[1+(-1)^{J_2}\right]
        (D'_1+D'^*_1+D'_2+D'^*_2)\nonumber\\
    &\qquad+\gamma^{J_3J_2}_{d'-}\left[1-(-1)^{J_2}\right]
        (D'_1-D'^*_1+D'_2-D'^*_2)\Big\}\nonumber\\
    &\qquad+\beta^{J_1J_2}_{d-}\Big\{\gamma^{J_3J_2}_{d'+}\left[1+(-1)^{J_2}\right]\nonumber\\
    &\qquad\times(D'_1-D'^*_1-D'_2+D'^*_2)\nonumber\\
    &\qquad+\gamma^{J_3J_2}_{d'-}\left[1-(-1)^{J_2}\right]
        (D'_1+D'^*_1-D'_2-D'^*_2)\Big\}\Bigg\}
\end{align}
where
\begin{eqnarray}
D'_1=D_1(\kappa=1)=D^{J_1}_{\mu d,0}D^{J_2*}_{\mu d,-d'}D^{J_3*}_{-d',0}
\end{eqnarray}
\noindent
and
\begin{eqnarray}
D'_2=D_2(\kappa=1)=D^{J_1*}_{\mu d,0}D^{J_2}_{\mu d,d'}D^{J_3}_{d',0}.
\end{eqnarray}
As $J_3$ can only take the values $0$ and $2$ in (\ref{W_mu}), we have defined
\begin{align}
&C^{J_3}=C^{J_3}_{\pm\frac{1}{2}}=\sqrt{3}\sum^{1}_{\alpha=0}(-1)^\alpha
\langle11;\alpha,-\alpha|J_3;0\rangle|C_\alpha|^2\\
&(J_3=0,2).\nonumber
\end{align}
The coefficients of the Wigner $D$-functions in (\ref{W_mu}) can be obtained from
\begin{align}\label{coeff W_mu}
&(-1)^{\frac{J_2}{2}(1+\mu)}B^{J_1}C^{J_3}\Bigg\{\beta^{J_1J_2}_{d+}
    \Big\{\gamma^{J_3J_2}_{d'+}\left[1+(-1)^{J_2}\right]\nonumber\\
    &\quad+\gamma^{J_2J_3}_{d'-}
    \left[1-(-1)^{J_2}\right](1-\delta_{d'0})\Big\}\nonumber\\
    &\quad+\beta^{J_1J_2}_{d-}\Big\{\gamma^{J_3J_2}_{d'+}\left[1+(-1)^{J_2}\right](1-\delta_{d0})\nonumber\\
        &\quad+\gamma^{J_3J_2}_{d'-}\left[1-(-1)^{J_2}\right](1-\delta_{d0})(1-\delta_{d'0})\Big\}\Bigg\}\nonumber\\
    &=2(2J_1+1)(2J_2+1)(2J_3+1)\nonumber\\
        &\quad\times\int W_\mu(\theta,\phi;\theta',\phi';\theta'',\phi'')D'^*_1d\Omega d\Omega' d\Omega''.
\end{align}

Similarly, if we only measure the polarization $\kappa$ of $\gamma_2$, the normalized combined angular distribution of $\gamma_1$, $\gamma_2$ and $e^-$ will become
\begin{align}\label{W_kappa}
&W_\kappa(\theta,\phi;\theta',\phi';\theta'',\phi'')\nonumber\\
&\quad=\frac{1}{4}\sum^{\pm1}_{\mu}\sum^{\pm\frac{1}{2}}_{\alpha_1}
    W_{\mu\kappa\alpha_1}(\theta,\phi;\theta',\phi';\theta'',\phi'')\nonumber\\
&\quad=\frac{1}{2(4\pi)^3}\sum^{0,2,4,6}_{J_1}B^{J_1}\sum^{0,2}_{J_3}C^{J_3}\sum^{4}_{J_2=0}
    (-1)^{\left(\frac{1-\kappa}{2}\right)(J_2+J_3)}\nonumber\\
    &\qquad\times\sum^{d_m}_{d=0}\sum^{d'_m}_{d'=0}
    \left(1-\frac{\delta_{d0}}{2}\right)\left(1-\frac{\delta_{d'0}}{2}\right)
    \Big\{\beta^{J_1J_2}_{d+}\nonumber\\
        &\qquad\times\left[1+(-1)^{J_2}\right]\left[\gamma^{J_3J_2}_{d'+}
    (D''_1+D''^*_1+D''_2+D''^*_2)\right.\nonumber\\
    &\qquad\left.+\gamma^{J_3J_2}_{d'-}(D''_1-D''^*_1+D''_2-D''^*_2)\right]
    -\beta^{J_1J_2}_{d-}\nonumber\\
        &\qquad\times\left[1-(-1)^{J_2}\right]\left[\gamma^{J_3J_2}_{d'+}
        (D''_1-D''^*_1-D''_2+D''^*_2)\right.\nonumber\\
        &\qquad\left.+\gamma^{J_3J_2}_{d'-}(D''_1+D''^*_1-D''_2-D''^*_2)\right]\Big\}
\end{align}
where
\begin{eqnarray}
D''_1=D_1(\mu=1)=D^{J_1}_{d,0}D^{J_2*}_{d,-\kappa d'}D^{J_3*}_{-\kappa d',0}
\end{eqnarray}
\noindent
and
\begin{eqnarray}
D''_2=D_2(\mu=1)=D^{J_1*}_{d,0}D^{J_2}_{d,\kappa d'}D^{J_3}_{\kappa d',0}.
\end{eqnarray}
The coefficients of the angular functions in (\ref{W_kappa}) can be obtained from
\begin{align}\label{coeff W_kappa}
&(-1)^{\left(\frac{1-\kappa}{2}\right)(J_2+J_3)}B^{J_1}C^{J_3}\Big\{\beta^{J_1J_2}_{d+}
    \left[1+(-1)^{J_2}\right]\nonumber\\
    &\quad\times\left[\gamma^{J_3J_2}_{d'+}
        +\gamma^{J_3J_2}_{d'-}(1-\delta_{d'0})\right]
    -\beta^{J_1J_2}_{d-}\left[1-(-1)^{J_2}\right]\nonumber\\
    &\quad\times\left[\gamma^{J_3J_2}_{d'+}(1-\delta_{d0})
        +\gamma^{J_3J_2}_{d'-}(1-\delta_{d0})(1-\delta_{d'0})\right]\Big\}\nonumber\\
&=2(2J_1+1)(2J_2+1)(2J_3+1)\nonumber\\
    &\quad\times\int W_\kappa(\theta,\phi;\theta',\phi';\theta'',\phi'')D''^*_1d\Omega d\Omega' d\Omega''
\end{align}
where again $J_3$ can only take the values $0$ and $2$.

If we are interested in only measuring the polarization $\alpha_1$ of $e^-$, the combined angular distribution of $\gamma_1$, $\gamma_2$ and $e^-$ will become
\begin{align}\label{W_alpha}
&W_{\alpha_1}(\theta,\phi;\theta',\phi';\theta'',\phi'')\nonumber\\
&\quad=\frac{1}{4}\sum^{\pm1}_{\mu}\sum^{\pm1}_{\kappa}
    W_{\mu\kappa\alpha_1}(\theta,\phi;\theta',\phi';\theta'',\phi'')\nonumber\\
&\quad=\frac{1}{4(4\pi)^3}\sum^{0,2,4,6}_{J_1}B^{J_1}
    \sum^{2}_{J_3=0}C^{J_3}_{\alpha_1}\sum^{4}_{J_2=0}\sum^{d_m}_{d=0}\sum^{d'_m}_{d'=0}
    \left(1-\frac{\delta_{d0}}{2}\right)\nonumber\\
    &\qquad\times\left(1-\frac{\delta_{d'0}}{2}\right)\Bigg\{\beta^{J_1J_2}_{d+}
    \left[1+(-1)^{J_2}\right]\Big\{\gamma^{J_3J_2}_{d'+}\nonumber\\
    &\qquad\times\left[1+(-1)^{J_3}\right]
    (D'''_1+D'''^*_1+D'''_2+D'''^*_2)+\gamma^{J_3J_2}_{d'-}\nonumber\\
    &\qquad\times\left[1-(-1)^{J_3}\right]
    (D'''_1-D'''^*_1+D'''_2-D'''^*_2)\Big\}\nonumber\\
    &\qquad+\beta^{J_1J_2}_{d-} \left[1-(-1)^{J_2}\right]\Big\{\gamma^{J_3J_2}_{d'+}\left[(-1)^{J_3}-1\right]\nonumber\\
    &\qquad\times(D'''_1-D'''^*_1-D'''_2+D'''^*_2)
    -\gamma^{J_3J_2}_{d'-}\left[1+(-1)^{J_3}\right]\nonumber\\
    &\qquad\times(D'''_1+D'''^*_1-D'''_2-D'''^*_2)\Big\}\Bigg\}
\end{align}
where
\begin{eqnarray}
D'''_1=D_1(\mu=\kappa=1)=D^{J_1}_{d,0}D^{J_2*}_{d,-d'}D^{J_3*}_{-d',0}
\end{eqnarray}
\noindent
and
\begin{eqnarray}
D'''_2=D_2(\mu=\kappa=1)=D^{J_1*}_{d,0}D^{J_2}_{d,d'}D^{J_3}_{d',0}.
\end{eqnarray}
The coefficients in (\ref{W_alpha}) can be obtained from
\begin{align}\label{coeff W_alpha}
&B^{J_1}C^{J_3}_{\alpha_1}\Bigg\{\beta^{J_1J_2}_{d+}\left[1+(-1)^{J_2}\right]
\Big\{\gamma^{J_3J_2}_{d'+}\left[1+(-1)^{J_3}\right]\nonumber\\
    &\quad+\gamma^{J_3J_2}_{d'-}\left[1-(-1)^{J_3}\right](1-\delta_{d'0})\Big\}
    +\beta^{J_1J_2}_{d-}\nonumber\\
        &\quad\times\left[1-(-1)^{J_2}\right]\Big\{\gamma^{J_3J_2}_{d'+}
    \left[(-1)^{J_3}-1\right](1-\delta_{d0})\nonumber\\
    &\quad-\gamma^{J_3J_2}_{d'-}\left[1+(-1)^{J_3}\right]
    (1-\delta_{d0})(1-\delta_{d'0})\Big\}\Bigg\}\nonumber\\
&=4(2J_1+1)(2J_2+1)(2J_3+1)\nonumber\\
    &\quad\times\int W_{\alpha_1}(\theta,\phi;\theta',\phi';\theta'',\phi'')D'''^*_1d\Omega d\Omega' d\Omega''.
\end{align}
Here, $J_3$ can take the values $0$, $1$ and $2$.

If we now average over the polarizations $\alpha_1$ of $e^-$ in (\ref{W_alpha}) as well, we get
\begin{align}\label{unpolarized}
&W(\theta,\phi;\theta',\phi';\theta'',\phi'')\nonumber\\
&\quad=\frac{1}{2}\sum^{\pm\frac{1}{2}}_{\alpha_1}
    W_{\alpha_1}(\theta,\phi;\theta',\phi';\theta'',\phi'')\nonumber\\
&\quad=\frac{1}{2(4\pi)^3}\sum^{0,2,4,6}_{J_1}B^{J_1}\sum^{0,2}_{J_3}C^{J_3}\sum^{4}_{J_2=0}
    \sum^{d_m}_{d=0}\sum^{d'_m}_{d'=0}\left(1-\frac{\delta_{d0}}{2}\right)\nonumber\\
    &\qquad\times\left(1-\frac{\delta_{d'0}}{2}\right)
    \left\{\beta^{J_1J_2}_{d+}\gamma^{J_3J_2}_{d'+}\left[1+(-1)^{J_2}\right]\right.\nonumber\\
    &\qquad\left.+\beta^{J_1J_2}_{d-}\gamma^{J_3J_2}_{d'-}\left[1-(-1)^{J_2}\right]\right\}\nonumber\\
    &\qquad\times\left[(-1)^{J_2}(D'''_1+D'''^*_1)+(D'''_2+D'''^*_2)\right]
\end{align}
Using (\ref{definition beta}) and (\ref{definition gamma}), we have
\begin{align}\label{simplify}
&\left(1-\frac{\delta_{d0}}{2}\right)\left(1-\frac{\delta_{d'0}}{2}\right)
    \Big\{\beta^{J_1J_2}_{d+}\gamma^{J_3J_2}_{d'+}\nonumber\\
    &\quad\times\left[1+(-1)^{J_2}\right]
    +\beta^{J_1J_2}_{d-}\gamma^{J_3J_2}_{d'-}\left[1-(-1)^{J_2}\right]\Big\}\nonumber\\
    &=\frac{\beta^{J_1J_2}_{d}\gamma^{J_3J_2}_{d'}}{2}
\end{align}
where
\begin{align}
\beta^{J_{1}J_{2}}_{d}&=\sqrt{35}\left(1-\frac{\delta_{d0}}{2}\right)\nonumber\\
    &\quad\times\sum_{s(d)}
        \left[A_{\frac{s+d}{2}}A^{*}_{\frac{s-d}{2}}
        +(-1)^{J_2}A^{*}_{\frac{s+d}{2}}A_{\frac{s-d}{2}}\right]\nonumber\\
        &\quad\times\left\langle22;\frac{s+d-4}{2},-\frac{s-d-4}{2}|J_2;d\right\rangle\nonumber\\
        &\quad\times\left\langle33;\frac{s+d-6}{2},-\frac{s-d-6}{2}|J_1;d\right\rangle\quad\\
\gamma^{J_{3}J_{2}}_{d'}&=\sqrt{15}\left(1-\frac{\delta_{d'0}}{2}\right)\nonumber\\
    &\quad\times\sum_{s'(d')}\left[E_{\frac{s'+d'}{2}}E^{*}_{\frac{s'-d'}{2}}
        +(-1)^{J_2}E^{*}_{\frac{s'+d'}{2}}E_{\frac{s'-d'}{2}}\right]\nonumber\\
        &\quad\times\left\langle22;\frac{s'+d'}{2},-\frac{s'-d'}{2}|J_2;d'\right\rangle\nonumber\\
        &\quad\times\left\langle11;\frac{s'+d'-2}{2},-\frac{s'-d'-2}{2}|J_3;d'\right\rangle
\end{align}
 By combining (\ref{simplify}) and (\ref{unpolarized}), we now recover our results in \cite{previous paper}, where the polarizations of the decay particles are not measured.

Using (\ref{coeff W_mu}), (\ref{coeff W_kappa}) or (\ref{coeff W_alpha}) it can be seen that once the combined angular distribution
         $W_{\mu}(\theta,\phi;\theta',\phi';\theta'',\phi'')$
,
         $W_{\kappa}(\theta,\phi;\theta',\phi';\theta'',\phi'')$
or
         $W_{\alpha_1}(\theta,\phi;\theta',\phi';\theta'',\phi'')$
is measured, one can also get the same information on the helicity amplitudes as one obtained from measuring the angular distribution function $W_{\mu\kappa\alpha_1}(\theta,\phi;\theta',\phi';\theta'',\phi'')$
where the polarizations of the three particles $\gamma_1$, $\gamma_2$ and $e^-$ are observed. In other words, by measuring the combined angular distribution of the decay particles $\gamma_1$, $\gamma_2$ and $e^-$ and the polarization of any one particle, we can get complete information on the helicity amplitudes in the radiative decay processes $\fourIdx{3}{}{}{3}{D}\rightarrow\fourIdx{3}{}{}{2}{P}+\gamma_1$ and $\fourIdx{3}{}{}{2}{P}\rightarrow\psi+\gamma_2$. In addition, we can also get the relative magnitudes of the helicity amplitudes in the production process $\bar{p}p\rightarrow\fourIdx{3}{}{}{3}{D}$ and in the final decay process $\psi\rightarrow e^+e^-$.
\section{Partially integrated angular distributions}
\label{sec:2}
The partially integrated angular distributions obtained from (\ref{wigner final_pol.}) will look a lot simpler and we will gain greater insight from them. There are three different cases in which the polarization and the angular distribution of only one particle ($\gamma_1$, $\gamma_2$ or $e^-$) are measured. We find that these results are identical to the single-particle angular distribution functions given in \cite{previous paper}, where the polarizations of the individual particles are not measured. So including the measurement of the polarizations in the single-particle angular distributions does not give us any further information. However, we will find that the measurement of the polarizations of the decay particles can provide us more information on the helicity amplitudes when we measure the simultaneous angular distributions of two particles. We now consider three different cases of two-particle angular distributions. We will express our results in terms of the spherical harmonics by using the following relation:
\begin{align}
D^J_{M0}=\sqrt{\frac{4\pi}{2J+1}}Y^*_{JM}
\end{align}

\textbf{Case 1}. We integrate over the angles $(\theta'',\phi'')$ or the direction of $e^-$ and then average over the polarization $\alpha_1$ of $e^-$. The combined angular distribution of $\gamma_1$ and $\gamma_2$, and the polarization of only one of the two particles are measured. The explicit expressions are given in the following.

Only $\mu$ is measured:
\begin{align}\label{case1 W_mu}
&\widetilde{\widetilde{W}}_{\mu}(\theta,\phi;\theta',\phi')\nonumber\\
&=\frac{1}{4}\sum^{\pm1}_{\kappa}\sum^{\pm\frac{1}{2}}_{\alpha_1}\int W_{\mu\kappa\alpha_1}(\theta,\phi;\theta',\phi';\theta'',\phi'')d\Omega''\nonumber\\
&=-\frac{1}{4\pi}\Bigg\{-\frac{1}{4\pi}
+\frac{1}{\sqrt{5}}B^0\beta^{02}_{0+}\gamma^{02}_{0+}Y_{00}(\theta,\phi)Y_{20}(\theta',\phi')\nonumber\\
&\quad+\frac{1}{3}B^0\beta^{04}_{0+}\gamma^{04}_{0+}Y_{00}(\theta,\phi)Y_{40}(\theta',\phi')\nonumber\\
&\quad+\frac{1}{\sqrt{5}}B^2\beta^{20}_{0+}\gamma^{00}_{0+}Y_{20}(\theta,\phi)Y_{00}(\theta',\phi')\nonumber\\
&\quad+\frac{1}{5}B^2\beta^{22}_{0+}\gamma^{02}_{0+}Y_{20}(\theta,\phi)Y_{20}(\theta',\phi')\nonumber\\
&\quad+\frac{2}{5}\left[B^2\beta^{22}_{1+}\gamma^{02}_{0+}
    \operatorname{Re}(Y_{2,\mu}(\theta,\phi)Y^*_{2,\mu}(\theta',\phi'))\right.\nonumber\\
&\quad -\operatorname{i}B^2\beta^{22}_{1-}\gamma^{02}_{0+}
        \operatorname{Im}(Y_{2,\mu}(\theta,\phi)Y^*_{2,\mu}(\theta',\phi'))\nonumber\\
&\quad+B^2\beta^{22}_{2+}\gamma^{02}_{0+}
        \operatorname{Re}(Y_{2,2\mu}(\theta,\phi)Y^*_{2,2\mu}(\theta',\phi'))\nonumber\\
&\quad-\left.\operatorname{i}B^2\beta^{22}_{2-}\gamma^{02}_{0+}
            \operatorname{Im}(Y_{2,2\mu}(\theta,\phi)Y^*_{2,2\mu}(\theta',\phi'))\right]\nonumber\\
&\quad+\frac{1}{3\sqrt{5}}B^2\beta^{24}_{0+}\gamma^{04}_{0+}Y_{20}(\theta,\phi)Y_{40}(\theta',\phi')\nonumber\\
&\quad+\frac{2}{3\sqrt{5}}\left[B^2\beta^{24}_{1+}\gamma^{04}_{0+}
    \operatorname{Re}(Y_{2,\mu}(\theta,\phi)Y^*_{4,\mu}(\theta',\phi'))\right.\nonumber\\
&\quad-\operatorname{i}B^2\beta^{24}_{1-}\gamma^{04}_{0+}
        \operatorname{Im}(Y_{2,\mu}(\theta,\phi)Y^*_{4,\mu}(\theta',\phi'))\nonumber\\
&\quad+B^2\beta^{24}_{2+}\gamma^{04}_{0+}
        \operatorname{Re}(Y_{2,2\mu}(\theta,\phi)Y^*_{4,2\mu}(\theta',\phi'))\nonumber\\
&\quad-\left.\operatorname{i}B^2\beta^{24}_{2-}\gamma^{04}_{0+}
            \operatorname{Im}(Y_{2,2\mu}(\theta,\phi)Y^*_{4,2\mu}(\theta',\phi'))\right]\nonumber\\
&\quad+\frac{1}{3}B^4\beta^{40}_{0+}\gamma^{00}_{0+}Y_{40}(\theta,\phi)Y_{00}(\theta',\phi')\nonumber\\
&\quad+\frac{1}{3\sqrt{5}}B^4\beta^{42}_{0+}\gamma^{02}_{0+}Y_{40}(\theta,\phi)Y_{20}(\theta',\phi')\nonumber\\
&\quad+\frac{2}{3\sqrt{5}}\left[B^4\beta^{42}_{1+}\gamma^{02}_{0+}
    \operatorname{Re}(Y_{4,\mu}(\theta,\phi)Y^*_{2,\mu}(\theta',\phi'))\right.\nonumber\\
&\quad-\operatorname{i}B^4\beta^{42}_{1-}\gamma^{02}_{0+}
        \operatorname{Im}(Y_{4,\mu}(\theta,\phi)Y^*_{2,\mu}(\theta',\phi'))\nonumber\\
&\quad+B^4\beta^{42}_{2+}\gamma^{02}_{0+}
        \operatorname{Re}(Y_{4,2\mu}(\theta,\phi)Y^*_{2,2\mu}(\theta',\phi'))\nonumber\\
&\quad-\left.\operatorname{i}B^4\beta^{42}_{2-}\gamma^{02}_{0+}
        \operatorname{Im}(Y_{4,2\mu}(\theta,\phi)Y^*_{2,2\mu}(\theta',\phi'))\right]\nonumber\\
&\quad+\frac{1}{9}B^4\beta^{44}_{0+}\gamma^{04}_{0+}Y_{40}(\theta,\phi)Y_{40}(\theta',\phi')\nonumber\\
&\quad+\frac{2}{9}\left[B^4\beta^{44}_{1+}\gamma^{04}_{0+}
    \operatorname{Re}(Y_{4,\mu}(\theta,\phi)Y^*_{4,\mu}(\theta',\phi'))\right.\nonumber\\
&\quad-\operatorname{i}B^4\beta^{44}_{1-}\gamma^{04}_{0+}
        \operatorname{Im}(Y_{4,\mu}(\theta,\phi)Y^*_{4,\mu}(\theta',\phi'))\nonumber\\
&\quad+B^4\beta^{44}_{2+}\gamma^{04}_{0+}
        \operatorname{Re}(Y_{4,2\mu}(\theta,\phi)Y^*_{4,2\mu}(\theta',\phi'))\nonumber\\
&\quad-\operatorname{i}B^4\beta^{44}_{2-}\gamma^{04}_{0+}
        \operatorname{Im}(Y_{4,2\mu}(\theta,\phi)Y^*_{4,2\mu}(\theta',\phi'))\nonumber\\
&\quad+B^4\beta^{44}_{3+}\gamma^{04}_{0+}
        \operatorname{Re}(Y_{4,3\mu}(\theta,\phi)Y^*_{4,3\mu}(\theta',\phi'))\nonumber\\
&\quad-\operatorname{i}B^4\beta^{44}_{3-}\gamma^{04}_{0+}
        \operatorname{Im}(Y_{4,3\mu}(\theta,\phi)Y^*_{4,3\mu}(\theta',\phi'))\nonumber\\
&\quad+B^4\beta^{44}_{4+}\gamma^{04}_{0+}
        \operatorname{Re}(Y_{4,4\mu}(\theta,\phi)Y^*_{4,4\mu}(\theta',\phi'))\nonumber\\
&\quad-\left.\operatorname{i}B^4\beta^{44}_{4-}\gamma^{04}_{0+}
        \operatorname{Im}(Y_{4,4\mu}(\theta,\phi)Y^*_{4,4\mu}(\theta',\phi'))\right]\nonumber\\
&\quad+\frac{1}{\sqrt{13}}B^6\beta^{60}_{0+}\gamma^{00}_{0+}Y_{60}(\theta,\phi)Y_{00}(\theta',\phi')\nonumber\\
&\quad+\frac{1}{\sqrt{65}}B^6\beta^{62}_{0+}\gamma^{02}_{0+}Y_{60}(\theta,\phi)Y_{20}(\theta',\phi')\nonumber\\
&\quad+\frac{2}{\sqrt{65}}\left[B^6\beta^{62}_{1+}\gamma^{02}_{0+}
    \operatorname{Re}(Y_{6,\mu}(\theta,\phi)Y^*_{2,\mu}(\theta',\phi'))\right.\nonumber\\
&\quad-\operatorname{i}B^6\beta^{62}_{1-}\gamma^{02}_{0+}
        \operatorname{Im}(Y_{6,\mu}(\theta,\phi)Y^*_{2,\mu}(\theta',\phi'))\nonumber\\
&\quad+B^6\beta^{62}_{2+}\gamma^{02}_{0+}
        \operatorname{Re}(Y_{6,2\mu}(\theta,\phi)Y^*_{2,2\mu}(\theta',\phi'))\nonumber\\
&\quad-\left.\operatorname{i}B^6\beta^{62}_{2-}\gamma^{02}_{0+}
        \operatorname{Im}(Y_{6,2\mu}(\theta,\phi)Y^*_{2,2\mu}(\theta',\phi'))\right]\nonumber\\
&\quad+\frac{1}{3\sqrt{13}}B^6\beta^{64}_{0+}\gamma^{04}_{0+}Y_{60}(\theta,\phi)Y_{40}(\theta',\phi')\nonumber\\
&\quad+\frac{2}{3\sqrt{13}}\left[B^6\beta^{64}_{1+}\gamma^{04}_{0+}
    \operatorname{Re}(Y_{6,\mu}(\theta,\phi)Y^*_{4,\mu}(\theta',\phi'))\right.\nonumber\\
&\quad-\operatorname{i}B^6\beta^{64}_{1-}\gamma^{04}_{0+}
        \operatorname{Im}(Y_{6,\mu}(\theta,\phi)Y^*_{4,\mu}(\theta',\phi'))\nonumber\\
&\quad+B^6\beta^{64}_{2+}\gamma^{04}_{0+}
        \operatorname{Re}(Y_{6,2\mu}(\theta,\phi)Y^*_{4,2\mu}(\theta',\phi'))\nonumber\\
&\quad-\operatorname{i}B^6\beta^{64}_{2-}\gamma^{04}_{0+}
        \operatorname{Im}(Y_{6,2\mu}(\theta,\phi)Y^*_{4,2\mu}(\theta',\phi'))\nonumber\\
&\quad+B^6\beta^{64}_{3+}\gamma^{04}_{0+}
        \operatorname{Re}(Y_{6,3\mu}(\theta,\phi)Y^*_{4,3\mu}(\theta',\phi'))\nonumber\\
&\quad-\operatorname{i}B^6\beta^{64}_{3-}\gamma^{04}_{0+}
        \operatorname{Im}(Y_{6,3\mu}(\theta,\phi)Y^*_{4,3\mu}(\theta',\phi'))\nonumber\\
&\quad+B^6\beta^{64}_{4+}\gamma^{04}_{0+}
        \operatorname{Re}(Y_{6,4\mu}(\theta,\phi)Y^*_{4,4\mu}(\theta',\phi'))\nonumber\\
&\quad-\left.\operatorname{i}B^6\beta^{64}_{4-}\gamma^{04}_{0+}
        \operatorname{Im}(Y_{6,4\mu}(\theta,\phi)Y^*_{4,4\mu}(\theta',\phi'))\right]\Bigg\}.
\end{align}

Only $\kappa$ is measured:
\begin{align}\label{Case1 W_kappa}
&\widetilde{\widetilde{W}}_{\kappa}(\theta,\phi;\theta',\phi')\nonumber\\
&=\frac{1}{4}\sum^{\pm1}_{\mu}\sum^{\pm\frac{1}{2}}_{\alpha_1}
    \int W_{\mu\kappa\alpha_1}(\theta,\phi;\theta',\phi';\theta'',\phi'')d\Omega''\nonumber\\
&=-\frac{1}{4\pi}\Bigg\{-\frac{1}{4\pi}
+\frac{1}{\sqrt{5}}B^0\beta^{02}_{0+}\gamma^{02}_{0+}Y_{00}(\theta,\phi)Y_{20}(\theta',\phi')\nonumber\\
&\quad+\frac{1}{3}B^0\beta^{04}_{0+}\gamma^{04}_{0+}Y_{00}(\theta,\phi)Y_{40}(\theta',\phi')\nonumber\\
&\quad+\frac{1}{\sqrt{5}}B^2\beta^{20}_{0+}\gamma^{00}_{0+}Y_{20}(\theta,\phi)Y_{00}(\theta',\phi')\nonumber\\
&\quad+\frac{2\operatorname{i}(-1)^{\frac{1}{2}(1-\kappa)}}{\sqrt{15}}B^2\beta^{21}_{1-}\gamma^{01}_{0+}
    \operatorname{Im}(Y_{21}(\theta,\phi)Y^*_{11}(\theta',\phi'))\nonumber\\
&\quad+\frac{1}{5}B^2\beta^{22}_{0+}\gamma^{02}_{0+}Y_{20}(\theta,\phi)Y_{20}(\theta',\phi')\nonumber\\
&\quad+\frac{2}{5}\left[B^2\beta^{22}_{1+}\gamma^{02}_{0+}
    \operatorname{Re}(Y_{21}(\theta,\phi)Y^*_{21}(\theta',\phi'))\right.\nonumber\\
&\quad+\left.B^2\beta^{22}_{2+}\gamma^{02}_{0+}
        \operatorname{Re}(Y_{22}(\theta,\phi)Y^*_{22}(\theta',\phi'))\right]\nonumber\\
&\quad+\frac{2\operatorname{i}(-1)^{\frac{1}{2}(1-\kappa)}}{\sqrt{35}}
    \left[B^2\beta^{23}_{1-}\gamma^{03}_{0+}
        \operatorname{Im}(Y_{21}(\theta,\phi)Y^*_{31}(\theta',\phi'))\right.\nonumber\\
&\quad+\left.B^2\beta^{23}_{2-}\gamma^{03}_{0+}
        \operatorname{Im}(Y_{22}(\theta,\phi)Y^*_{23}(\theta',\phi'))\right]\nonumber\\
&\quad+\frac{1}{3\sqrt{5}}B^2\beta^{24}_{0+}\gamma^{04}_{0+}Y_{20}(\theta,\phi)Y_{40}(\theta',\phi')\nonumber\\
&\quad+\frac{2}{3\sqrt{5}}\left[B^2\beta^{24}_{1+}\gamma^{04}_{0+}
    \operatorname{Re}(Y_{21}(\theta,\phi)Y^*_{41}(\theta',\phi'))\right.\nonumber\\
&\quad+\left.B^2\beta^{24}_{2+}\gamma^{04}_{0+}
        \operatorname{Re}(Y_{22}(\theta,\phi)Y^*_{42}(\theta',\phi'))\right]\nonumber\\
&\quad+\frac{1}{3}B^4\beta^{40}_{0+}\gamma^{00}_{0+}Y_{40}(\theta,\phi)Y_{00}(\theta',\phi')\nonumber\\
&\quad+\frac{2\operatorname{i}(-1)^{\frac{1}{2}(1-\kappa)}}{3\sqrt{3}}B^4\beta^{41}_{1-}\gamma^{01}_{0+}
    \operatorname{Im}(Y_{41}(\theta,\phi)Y^*_{11}(\theta',\phi'))\nonumber\\
&\quad+\frac{1}{3\sqrt{5}}B^4\beta^{42}_{0+}\gamma^{02}_{0+}Y_{40}(\theta,\phi)Y_{20}(\theta',\phi')\nonumber\\
&\quad+\frac{2}{3\sqrt{5}}\left[B^4\beta^{42}_{1+}\gamma^{02}_{0+}
    \operatorname{Re}(Y_{41}(\theta,\phi)Y^*_{21}(\theta',\phi'))\right.\nonumber\\
&\quad+\left.B^4\beta^{42}_{2+}\gamma^{02}_{0+}
        \operatorname{Re}(Y_{42}(\theta,\phi)Y^*_{22}(\theta',\phi'))\right]\nonumber\\
&\quad+\frac{2\operatorname{i}(-1)^{\frac{1}{2}(1-\kappa)}}{3\sqrt{7}}
    \left[B^4\beta^{43}_{1-}\gamma^{03}_{0+}\operatorname{Im}(Y_{41}(\theta,\phi)Y^*_{31}(\theta',\phi'))\right.\nonumber\\
&\quad+B^4\beta^{43}_{2-}\gamma^{03}_{0+}\operatorname{Im}(Y_{42}(\theta,\phi)Y^*_{32}(\theta',\phi'))\nonumber\\
&\quad+\left.B^4\beta^{43}_{3-}\gamma^{03}_{0+}
        \operatorname{Im}(Y_{43}(\theta,\phi)Y^*_{33}(\theta',\phi'))\right]\nonumber\\
&\quad+\frac{1}{9}B^4\beta^{44}_{0+}\gamma^{04}_{0+}Y_{40}(\theta,\phi)Y_{40}(\theta',\phi')\nonumber\\
&\quad+\frac{2}{9}\left[B^4\beta^{44}_{1+}\gamma^{04}_{0+}
    \operatorname{Re}(Y_{41}(\theta,\phi)Y^*_{41}(\theta',\phi'))\right.\nonumber\\
&\quad+B^4\beta^{44}_{2+}\gamma^{04}_{0+}
        \operatorname{Re}(Y_{42}(\theta,\phi)Y^*_{42}(\theta',\phi'))\nonumber\\
&\quad+B^4\beta^{44}_{3+}\gamma^{04}_{0+}
        \operatorname{Re}(Y_{43}(\theta,\phi)Y^*_{43}(\theta',\phi'))\nonumber\\
&\quad+\left.B^4\beta^{44}_{4+}\gamma^{04}_{0+}
        \operatorname{Re}(Y_{44}(\theta,\phi)Y^*_{44}(\theta',\phi'))\right]\nonumber\\
&\quad+\frac{1}{\sqrt{13}}B^6\beta^{60}_{0+}\gamma^{00}_{0+}Y_{60}(\theta,\phi)Y_{00}(\theta',\phi')\nonumber\\
&\quad+\frac{2\operatorname{i}(-1)^{\frac{1}{2}(1-\kappa)}}{\sqrt{39}}B^6\beta^{61}_{1-}\gamma^{01}_{0+}
    \operatorname{Im}(Y_{61}(\theta,\phi)Y^*_{11}(\theta',\phi'))\nonumber\\
&\quad+\frac{1}{\sqrt{65}}B^6\beta^{62}_{0+}\gamma^{02}_{0+}Y_{60}(\theta,\phi)Y_{20}(\theta',\phi')\nonumber\\
&\quad+\frac{2}{\sqrt{65}}\left[B^6\beta^{62}_{1+}\gamma^{02}_{0+}
    \operatorname{Re}(Y_{61}(\theta,\phi)Y^*_{21}(\theta',\phi'))\right.\nonumber\\
&\quad+\left.B^6\beta^{62}_{2+}\gamma^{02}_{0+}\operatorname{Re}(Y_{62}(\theta,\phi)Y^*_{22}(\theta',\phi'))\right]\nonumber\\
&\quad+\frac{2\operatorname{i}(-1)^{\frac{1}{2}(1-\kappa)}}{\sqrt{112}}
    \left[B^6\beta^{63}_{1-}\gamma^{03}_{0+}
        \operatorname{Im}(Y_{61}(\theta,\phi)Y^*_{31}(\theta',\phi'))\right.\nonumber\\
&\quad+B^6\beta^{63}_{2-}\gamma^{03}_{0+}
        \operatorname{Im}(Y_{62}(\theta,\phi)Y^*_{32}(\theta',\phi'))\nonumber\\
&\quad+\left.B^6\beta^{63}_{3-}\gamma^{03}_{0+}
       \operatorname{Im}(Y_{63}(\theta,\phi)Y^*_{33}(\theta',\phi'))\right]\nonumber\\
&\quad+\frac{1}{3\sqrt{13}}B^6\beta^{64}_{0+}\gamma^{04}_{0+}Y_{60}(\theta,\phi)Y_{40}(\theta',\phi')\nonumber\\
&\quad+\frac{2}{3\sqrt{13}}\left[B^6\beta^{64}_{1+}\gamma^{04}_{0+}
    \operatorname{Re}(Y_{61}(\theta,\phi)Y^*_{41}(\theta',\phi'))\right.\nonumber\\
&\quad+B^6\beta^{64}_{2+}\gamma^{04}_{0+}
        \operatorname{Re}(Y_{62}(\theta,\phi)Y^*_{42}(\theta',\phi'))\nonumber\\
&\quad+B^6\beta^{64}_{3+}\gamma^{04}_{0+}
        \operatorname{Re}(Y_{63}(\theta,\phi)Y^*_{43}(\theta',\phi'))\nonumber\\
&\quad+\left.B^6\beta^{64}_{4+}\gamma^{04}_{0+}
        \operatorname{Re}(Y_{64}(\theta,\phi)Y^*_{44}(\theta',\phi'))\right]\Bigg\}.
\end{align}
 An inspection of (\ref{case1 W_mu}) and (\ref{Case1 W_kappa}) shows that the magnitudes of the $A$, $B$ and $E$ helicity amplitudes as well as the cosines and sines of the relative phases of the $A$ helicity amplitudes can be extracted from the measurement of either $\widetilde{\widetilde{W}}_{\mu}$ or $\widetilde{\widetilde{W}}_{\kappa}$. It should be noted that the measurement of the polarization of one of the decay particles is essential for getting the sines of the relative phases among the $A$ helicity amplitudes. This is not possible for the unpolarized case.

\textbf{Case 2}. We integrate over $(\theta,\phi)$ or the direction of $\gamma_1$ and average over the polarization $\mu$ of $\gamma_1$. The combined angular distribution of $\gamma_2$ and $e^-$ and the polarization of either one of them are measured. The expressions are given in the following.

Only $\kappa$ is measured:
\begin{align}\label{case2 W_kappa}
&\widetilde{\widetilde{W}}_\kappa(\theta',\phi';\theta'',\phi'')\nonumber\\
&=\frac{1}{4}\sum^{\pm1}_{\mu}\sum^{\pm\frac{1}{2}}_{\alpha_1}\int
    W_{\mu\kappa\alpha_1}(\theta,\phi;\theta',\phi';\theta'',\phi'')d\Omega\nonumber\\
&=-\frac{1}{4\pi}\Bigg\{-\frac{1}{4\pi}
+\frac{1}{\sqrt{5}}C^2\beta^{00}_{0+}\gamma^{20}_{0+}Y_{00}(\theta',\phi')Y_{20}(\theta'',\phi'')\nonumber\\
&\quad+\frac{1}{\sqrt{5}}C^0\beta^{02}_{0+}\gamma^{02}_{0+}Y_{20}(\theta',\phi')Y_{00}(\theta'',\phi'')\nonumber\\
&\quad+\frac{1}{5}C^2\beta^{02}_{0+}\gamma^{22}_{0+}Y_{20}(\theta',\phi')Y_{20}(\theta'',\phi'')\nonumber\\
&\quad+\frac{2}{5}C^2\beta^{02}_{0+}
    \left[-\gamma^{22}_{1+}\operatorname{Re}(Y_{2,\kappa}(\theta',\phi')Y_{2,\kappa}(\theta'',\phi''))\right.\nonumber\\
&\quad+\operatorname{i}\gamma^{22}_{1-}
            \operatorname{Im}(Y_{2,\kappa}(\theta',\phi')Y_{2,\kappa}(\theta'',\phi''))\nonumber\\
&\quad+\gamma^{22}_{2+}\operatorname{Re}(Y_{2,2\kappa}(\theta',\phi')Y_{2,2\kappa}(\theta'',\phi''))\nonumber\\
&\quad-\left.\operatorname{i}\gamma^{22}_{2-}
            \operatorname{Im}(Y_{2,2\kappa}(\theta',\phi')Y_{2,2\kappa}(\theta'',\phi''))\right]\nonumber\\
&\quad+\frac{1}{3}C^0\beta^{04}_{0+}\gamma^{04}_{0+}Y_{40}(\theta',\phi')Y_{00}(\theta'',\phi'')\nonumber\\
&\quad+\frac{1}{3\sqrt{5}}C^2\beta^{04}_{0+}\gamma^{24}_{0+}Y_{40}(\theta',\phi')Y_{20}(\theta'',\phi'')\nonumber\\
&\quad+\frac{2}{3\sqrt{5}}C^2\beta^{04}_{0+}\left[-\gamma^{24}_{1+}
    \operatorname{Re}(Y_{4,\kappa}(\theta',\phi')Y_{2,\kappa}(\theta'',\phi''))\right.\nonumber\\
&\quad+\operatorname{i}\gamma^{24}_{1-}
        \operatorname{Im}(Y_{4,\kappa}(\theta',\phi')Y_{2,\kappa}(\theta'',\phi''))\nonumber\\
&\quad+\gamma^{24}_{2+}
        \operatorname{Re}(Y_{4,2\kappa}(\theta',\phi')Y_{2,2\kappa}(\theta'',\phi''))\nonumber\\
&\quad-\left.\operatorname{i}\gamma^{24}_{2-}
            \operatorname{Im}(Y_{4,2\kappa}(\theta',\phi')Y_{2,2\kappa}(\theta'',\phi''))\right]\Bigg\}.
\end{align}

Only $\alpha_1$ is measured:
\begin{align}\label{Case2 W_alpha_1}
&\widetilde{\widetilde{W}}_{\alpha_1}(\theta',\phi';\theta'',\phi'')\nonumber\\
&=\frac{1}{4}\sum^{\pm1}_{\mu}\sum^{\pm1}_{\kappa}\int
    W_{\mu\kappa\alpha_1}(\theta,\phi;\theta',\phi';\theta'',\phi'')d\Omega\nonumber\\
&=-\frac{1}{4\pi}\Bigg\{-\frac{1}{4\pi}
+\frac{1}{\sqrt{5}}C^2\beta^{00}_{0+}\gamma^{20}_{0+}Y_{00}(\theta',\phi')Y_{20}(\theta'',\phi'')\nonumber\\
&\quad+\frac{1}{\sqrt{5}}C^0\beta^{02}_{0+}\gamma^{02}_{0+}
    Y_{20}(\theta',\phi')Y_{00}(\theta'',\phi'')\nonumber\\
&\quad+\frac{2\operatorname{i}}{\sqrt{15}}C^1_{\alpha_1}\beta^{02}_{0+}\gamma^{12}_{1-}
    \operatorname{Im}(Y_{21}(\theta',\phi')Y_{11}(\theta'',\phi''))\nonumber\\
&\quad+\frac{1}{5}C^2\beta^{02}_{0+}\gamma^{22}_{0+}Y_{20}(\theta',\phi')Y_{20}(\theta'',\phi'')\nonumber\\
&\quad-\frac{2}{5}C^2\beta^{02}_{0+}\gamma^{22}_{1+}
    \operatorname{Re}(Y_{21}(\theta',\phi')Y_{21}(\theta'',\phi''))\nonumber\\
&\quad+\frac{2}{5}C^2\beta^{02}_{0+}\gamma^{22}_{2+}
    \operatorname{Re}(Y_{22}(\theta',\phi')Y_{22}(\theta'',\phi''))\nonumber\\
&\quad+\frac{1}{3}C^0\beta^{04}_{0+}\gamma^{04}_{0+}Y_{40}(\theta',\phi')Y_{00}(\theta'',\phi'')\nonumber\\
&\quad+\frac{2\operatorname{i}}{3\sqrt{3}}C^1_{\alpha_1}\beta^{04}_{0+}\gamma^{14}_{1-}
    \operatorname{Im}(Y_{41}(\theta',\phi')Y_{11}(\theta'',\phi''))\nonumber\\
&\quad+\frac{1}{3\sqrt{5}}C^2\beta^{04}_{0+}\gamma^{24}_{0+}Y_{40}(\theta',\phi')Y_{20}(\theta'',\phi'')\nonumber\\
&\quad-\frac{2}{3\sqrt{5}}C^2\beta^{04}_{0+}\gamma^{24}_{1+}
    \operatorname{Re}(Y_{41}(\theta',\phi')Y_{21}(\theta'',\phi''))\nonumber\\
&\quad+\frac{2}{3\sqrt{5}}C^2\beta^{04}_{0+}\gamma^{24}_{2+}
    \operatorname{Re}(Y_{42}(\theta',\phi')Y_{22}(\theta'',\phi''))\Bigg\}.
\end{align}
An examination of (\ref{case2 W_kappa}) and (\ref{Case2 W_alpha_1}) shows that we can obtain the magnitudes of the $E$ helicity amplitudes as well as both the cosines and sines of the relative phases of the $E$ helicity amplitudes when the simultaneous angular distribution of $\gamma_2$ and $e^-$ with the polarization of either one particles is measured. As in case $1$, the measurement of the polarization is essential for getting the sines of the relative phases of these helicity amplitudes uniquely. It is worth noting that we can now obtain all the information on the helicity amplitudes from the measurement of the joint angular distributions of only two particles.

\textbf{Case 3}. We integrate over $(\theta',\phi')$ or the direction of $\gamma_2$ and then average over the polarization $\kappa$ of $\gamma_2$. The combined angular distribution of $\gamma_1$ and $e^-$ and the polarization of either one of them are measured. Since we cannot obtain any useful information from this case, we do not provide the long expressions here.

\section{Concluding remarks}
\label{sec:3}
We have derived the model-independent expressions for the combined angular distribution of the final photons ($\gamma_1$ and $\gamma_2$) and electron ($e^-$) in the cascade process, $\bar{p}+p\rightarrow\fourIdx{3}{}{}{3}{D}\rightarrow\fourIdx{3}{}{}{2}{P}+\gamma_1\rightarrow
(\psi+\gamma_2)+\gamma_1\rightarrow(e^++e^-)+\gamma_1+\gamma_2$, when $\bar{p}$ and $p$ are unpolarized and the polarization of any one of the three decay particles is measured. Our expressions are based only on the general principles of quantum mechanics and the symmetry of the problem. We have also derived the partially integrated angular distribution functions which give the two-particle angular distributions of ($\gamma_1$, $\gamma_2$) and ($\gamma_2$, $e^-$) with the measurement of the polarization of one particle in each cases. Once these polarized angular distributions are experimentally measured, our expressions can be used to extract the information of all the independent helicity amplitudes in the radiative decay processes $\fourIdx{3}{}{}{3}{D}\rightarrow\fourIdx{3}{}{}{2}{P}+\gamma_1$ and $\fourIdx{3}{}{}{2}{P}\rightarrow\psi+\gamma_2$. In fact, the analysis of the angular correlations in the final decay products will serve to verify the presence of the intermediate $\fourIdx{3}{}{}{3}{D}$ state and its $J^{PC}$ quantum numbers in the cascade process. The experimentally determined values of the helicity amplitudes can then be compared with the predictions of various dynamical models.

The great advantage of measuring the angular distributions with the polarization of one particle is that one can obtain not only the relative magnitudes of the helicity amplitudes but also both the cosines and sines of the relative phases of the helicity amplitudes in the decay processes $\fourIdx{3}{}{}{3}{D}\rightarrow\fourIdx{3}{}{}{2}{P}+\gamma_1$ and $\fourIdx{3}{}{}{2}{P}\rightarrow\psi+\gamma_2$. This is important because the helicity amplitudes are in general complex \cite{KJ Sebastian}. Therefore by measuring the combined angular distribution of $\gamma_1$, $\gamma_2$ and $e^-$ with the polarization of any one of the three particles, we can obtain complete information on the helicity amplitudes in the two radiative decay processes. Alternatively, we can get the same information by measuring the two-particle angular distribution of $\gamma_2$ and $e^-$ and that of $\gamma_1$ and $\gamma_2$ with the polarization of either one of the two particles.

Both the theorists and the experimentalists would like to express their results in terms of the multipole amplitudes in the radiative transitions $\fourIdx{3}{}{}{3}{D}\rightarrow\fourIdx{3}{}{}{2}{P}+\gamma_1$ and $\fourIdx{3}{}{}{2}{P}\rightarrow\psi+\gamma_2$. The relationship between the helicity and the multipole amplitudes are given by the orthogonal transformations \cite{Mok 2009,phys rev D13 1203 (1976)}:
\begin{align}
&A_i=\sum^5_{k=1}a_k\sqrt{\frac{2k+1}{5}}
    \langle k,-1;3,(i-1)|2,(i-2)\rangle\label{multipole amp. A}\\
&(i=0,1,2,3,4),\nonumber
\end{align}
\noindent
and
\begin{align}
&E_j=\sum^3_{k=1}e_k\sqrt{\frac{2k+1}{5}}
    \langle k,1;1,(j-1)|2,j\rangle \label{multipole amp. E}\\
&(j=0,1,2),\nonumber
\end{align}
where $a_k$ and $e_k$ are the radiative multipole amplitudes in $\fourIdx{3}{}{}{3}{D}\rightarrow\fourIdx{3}{}{}{2}{P}+\gamma_1$ and $\fourIdx{3}{}{}{2}{P}\rightarrow\psi+\gamma_2$, respectively. Since the transformations of (\ref{multipole amp. A}) and (\ref{multipole amp. E}) are orthogonal,
\begin{align}
\sum^{4}_{i=0}|A_i|^2=\sum^{5}_{k=1}|a_k|^2=1\text{ ,}\quad
\sum^{2}_{j=0}|E_j|^2=\sum^{3}_{k=1}|e_k|^2=1\text{ .}
\end{align}
It is noteworthy that the decay process $\fourIdx{3}{}{}{3}{D}\rightarrow\fourIdx{3}{}{}{2}{P}+\gamma_1$ has five independent helicity amplitudes corresponding to five multipole amplitudes $E1$, $M2$, $E3$, $M4$ and $E5$. In any potential model for heavy quarkonia, the $M4$ or higher multipole amplitudes is zero to order $v^2/c^2$ because in this approximation there is no fourth or higher rank tensor component in the transition operator \cite{phys rev D55 225 (1997)}. So by measuring the angular distributions, one can further test the validity of the non-relativistic potential models.

\appendix
\section{Expressions of coefficients}
\label{sec:apdix}
\subsection{Expressions of $B^{J_1}$}
\label{sec:apdix subsec1}
\begin{align}
    B^0&=-1\\
    B^2&=\frac{2}{\sqrt{3}}\left(|B_0|^2+\frac{3}{4}|B_1|^2\right)\\
    B^4&=-3\sqrt{\frac{2}{11}}\left(|B_0|^2+\frac{1}{6}|B_1|^2\right)\\
    B^6&=\frac{10}{\sqrt{33}}\left(|B_0|^2-\frac{3}{4}|B_1|^2\right)
\end{align}

\subsection{Expressions of $C^{J_3}_{\alpha_1}$}
\label{sec:apdix subsec2}
\begin{align}
   C^0_{\pm\frac{1}{2}}&=C^0=-1\\
   C^1_{\pm\frac{1}{2}}&=\pm\sqrt{\frac{3}{2}}|C_1|^2\\
   C^2_{\pm\frac{1}{2}}&=C^2=\sqrt{2}\left(|C_0|^2-\frac{1}{2}|C_1|^2\right)
\end{align}
\subsection{Expressions of $\beta^{J_1J_2}_{d\pm}$}
\label{sec:apdix subsec3}
\begin{align}
   \beta^{00}_{0+}&=1\\
   \beta^{20}_{0+}&=\frac{5}{2\sqrt{3}}
    \left(|A_0|^2-\frac{3}{5}|A_2|^2-\frac{4}{5}|A_3|^2-\frac{3}{5}|A_4|^2\right)\\
   \beta^{40}_{0+}&=\frac{3}{\sqrt{22}}\left(|A_0|^2-\frac{7}{3}|A_1|^2\right.\nonumber\\
        &\quad\left.+\frac{1}{3}|A_2|^2+2|A_3|^2+\frac{1}{3}|A_4|^2\right)\\
   \beta^{60}_{0+}&=\frac{1}{2\sqrt{33}}(|A_0|^2-6|A_1|^2\nonumber\\
        &\quad+15|A_2|^2-20|A_3|^2+15|A_4|^2)\\
   \beta^{01}_{0+}&=-\sqrt{2}\left(|A_0|^2+\frac{1}{2}|A_1|^2-\frac{1}{2}|A_3|^2-|A_4|^2\right)\\
   \beta^{21}_{0+}&=-\frac{5}{\sqrt{6}}\left(|A_0|^2+\frac{2}{5}|A_3|^2+\frac{3}{5}|A_4|^2\right)\\
   \beta^{41}_{0+}&=-\frac{3}{\sqrt{11}}
    \left(|A_0|^2-\frac{7}{6}|A_1|^2-|A_3|^2-\frac{1}{3}|A_4|^2\right)\\
   \beta^{61}_{0+}&=-\frac{1}{\sqrt{66}}(|A_0|^2-3|A_1|^2+10|A_3|^2-15|A_4|^2)\\
   \beta^{02}_{0+}&=\sqrt{\frac{10}{7}}
    \left(|A_0|^2-\frac{1}{2}|A_1|^2-|A_2|^2-\frac{1}{2}|A_3|^2+|A_4|^2\right)\\
   \beta^{22}_{0+}&=5\sqrt{\frac{5}{42}}
    \left(|A_0|^2+\frac{3}{5}|A_2|^2+\frac{2}{5}|A_3|^2-\frac{3}{5}|A_4|^2\right)\\
   \beta^{42}_{0+}&=3\sqrt{\frac{5}{77}}\left(|A_0|^2+\frac{7}{6}|A_1|^2\right.\nonumber\\
        &\quad\left.-\frac{1}{3}|A_2|^2-|A_3|^2+\frac{1}{3}|A_4|^2\right)\\
   \beta^{62}_{0+}&=\sqrt{\frac{5}{462}}(|A_0|^2+3|A_1|^2\nonumber\\
        &\quad-15|A_2|^2+10|A_3|^2+15|A_4|^2)\\
   \beta^{03}_{0+}&=-\frac{1}{\sqrt{2}}(|A_0|^2-2|A_1|^2+2|A_3|^2-|A_4|^2)\\
   \beta^{23}_{0+}&=-\frac{5}{2\sqrt{6}}\left(|A_0|^2-\frac{8}{5}|A_3|^2+\frac{3}{5}|A_4|^2\right)\\
   \beta^{43}_{0+}&=-\frac{3}{2\sqrt{11}}
    \left(|A_0|^2+\frac{14}{3}|A_1|^2+4|A_3|^2-\frac{1}{3}|A_4|^2\right)\\
   \beta^{63}_{0+}&=-\frac{1}{2\sqrt{66}}(|A_0|^2+12|A_1|^2-40|A_3|^2-15|A_4|^2)\\
   \beta^{04}_{0+}&=\frac{1}{\sqrt{14}}(|A_0|^2-4|A_1|^2+6|A_2|^2-4|A_3|^2+|A_4|^2)\\
   \beta^{24}_{0+}&=\frac{5}{2\sqrt{42}}
    \left(|A_0|^2-\frac{18}{5}|A_2|^2+\frac{16}{5}|A_3|^2-\frac{3}{5}|A_4|^2\right)\\
   \beta^{44}_{0+}&=\frac{3}{2\sqrt{77}}\left(|A_0|^2+\frac{28}{3}|A_1|^2\right.\nonumber\\
        &\quad+\left.2|A_2|^2-8|A_3|^2+\frac{1}{3}|A_4|^2\right)\\
   \beta^{64}_{0+}&=\frac{1}{2\sqrt{462}}(|A_0|^2+24|A_1|^2\nonumber\\
        &\quad+90|A_2|^2+80|A_3|^2+15|A_4|^2)\\
   \beta^{21}_{1+}&=-\frac{5}{2\sqrt{3}}\left[\operatorname{Re}(A_1A^*_0)+\frac{3}{\sqrt{10}}\operatorname{Re}(A_2A^*_1)\right.\nonumber\\
        &\quad\left.+\frac{\sqrt{3}}{5}\operatorname{Re}(A_3A^*_2)-\frac{\sqrt{2}}{5}\operatorname{Re}(A_4A^*_3)\right]\\
   \beta^{41}_{1+}&=-\sqrt{\frac{15}{11}}\left[\operatorname{Re}(A_1A^*_0)-\frac{4}{\sqrt{10}}\operatorname{Re}(A_2A^*_1)\right.\nonumber\\
        &\quad\left.-\frac{\sqrt{3}}{2}\operatorname{Re}(A_3A^*_2)+\frac{1}{\sqrt{2}}\operatorname{Re}(A_4A^*_3)\right]\\
   \beta^{61}_{1+}&=-\frac{1}{2}\sqrt{\frac{7}{33}}\left[\operatorname{Re}(A_1A^*_0)-6\sqrt{\frac{5}{2}}\operatorname{Re}(A_2A^*_1)\right.\nonumber\\
        &\quad\left.+5\sqrt{3}\operatorname{Re}(A_3A^*_2)-5\sqrt{2}\operatorname{Re}(A_4A^*_3)\right]\\
   \beta^{22}_{1+}&=\frac{5}{2}\sqrt{\frac{5}{7}}\left[\operatorname{Re}(A_1A^*_0)+\frac{1}{\sqrt{10}}\operatorname{Re}(A_2A^*_1)\right.\nonumber\\
        &\quad\left.-\frac{1}{5\sqrt{3}}\operatorname{Re}(A_3A^*_2)+\frac{\sqrt{2}}{5}\operatorname{Re}(A_4A^*_3)\right]\\
   \beta^{42}_{1+}&=\frac{15}{\sqrt{77}}\left[\operatorname{Re}(A_1A^*_0)-\frac{4}{3\sqrt{10}}\operatorname{Re}(A_2A^*_1)\right.\nonumber\\
        &\quad\left.+\frac{1}{2\sqrt{3}}\operatorname{Re}(A_3A^*_2)-\frac{1}{\sqrt{2}}\operatorname{Re}(A_4A^*_3)\right]\\
   \beta^{62}_{1+}&=\frac{1}{2}\sqrt{\frac{5}{11}}\left[\operatorname{Re}(A_1A^*_0)-\sqrt{\frac{5}{2}}\operatorname{Re}(A_2A^*_1)\right.\nonumber\\
        &\quad\left.-\frac{5}{\sqrt{3}}\operatorname{Re}(A_3A^*_2)+5\sqrt{2}\operatorname{Re}(A_4A^*_3)\right]\\
   \beta^{23}_{1+}&=-\frac{5}{2\sqrt{2}}\left[\operatorname{Re}(A_1A^*_0)-\sqrt{\frac{2}{5}}\operatorname{Re}(A_2A^*_1)\right.\nonumber\\
        &\quad\left.-\frac{2}{5\sqrt{3}}\operatorname{Re}(A_3A^*_2)-\frac{\sqrt{2}}{5}\operatorname{Re}(A_4A^*_3)\right]\\
   \beta^{43}_{1+}&=-\frac{3}{2}\sqrt{\frac{10}{11}}\left[\operatorname{Re}(A_1A^*_0)
    +\frac{4}{3}\sqrt{\frac{2}{5}}\operatorname{Re}(A_2A^*_1)\right.\nonumber\\
        &\quad\left.+\frac{1}{\sqrt{3}}\operatorname{Re}(A_3A^*_2)
    +\frac{1}{\sqrt{2}}\operatorname{Re}(A_4A^*_3)\right]\\
   \beta^{63}_{1+}&=-\frac{1}{2}\sqrt{\frac{7}{22}}\left[\operatorname{Re}(A_1A^*_0)+\sqrt{10}\operatorname{Re}(A_2A^*_1)\right.\nonumber\\
        &\quad\left.-\frac{10}{\sqrt{3}}\operatorname{Re}(A_3A^*_2)-5\sqrt{2}\operatorname{Re}(A_4A^*_3)\right]\\
   \beta^{24}_{1+}&=\frac{5}{2}\sqrt{\frac{5}{42}}\left[\operatorname{Re}(A_1A^*_0)-3\sqrt{\frac{2}{5}}\operatorname{Re}(A_2A^*_1)\right.\nonumber\\
        &\quad\left.+\frac{2\sqrt{3}}{5}\operatorname{Re}(A_3A^*_2)+\frac{\sqrt{2}}{5}\operatorname{Re}(A_4A^*_3)\right]\\
   \beta^{44}_{1+}&=\frac{5}{2}\sqrt{\frac{6}{7}}\left[\operatorname{Re}(A_1A^*_0)+4\sqrt{\frac{2}{5}}\operatorname{Re}(A_2A^*_1)\right.\nonumber\\
        &\quad\left.-\sqrt{3}\operatorname{Re}(A_3A^*_2)-\frac{1}{\sqrt{2}}\operatorname{Re}(A_4A^*_3)\right]\\
   \beta^{64}_{1+}&=\frac{1}{2}\sqrt{\frac{5}{66}}\left[\operatorname{Re}(A_1A^*_0)+3\sqrt{10}\operatorname{Re}(A_2A^*_1)\right.\nonumber\\
        &\quad\left.+10\sqrt{3}\operatorname{Re}(A_3A^*_2)+5\sqrt{2}\operatorname{Re}(A_4A^*_3)\right]\\
   \beta^{21}_{1-}&=-\frac{5\operatorname{i}}{2\sqrt{3}}\left[\operatorname{Im}(A_1A^*_0)+\frac{3}{\sqrt{10}}\operatorname{Im}(A_2A^*_1)\right.\nonumber\\
        &\quad\left.+\frac{\sqrt{3}}{5}\operatorname{Im}(A_3A^*_2)-\frac{\sqrt{2}}{5}\operatorname{Im}(A_4A^*_3)\right]\\
   \beta^{41}_{1-}&=-\operatorname{i}\sqrt{\frac{15}{11}}\left[\operatorname{Im}(A_1A^*_0)-\frac{4}{\sqrt{10}}\operatorname{Im}(A_2A^*_1)\right.\nonumber\\
        &\quad\left.-\frac{\sqrt{3}}{2}\operatorname{Im}(A_3A^*_2)+\frac{1}{\sqrt{2}}\operatorname{Im}(A_4A^*_3)\right]\\
   \beta^{61}_{1-}&=-\frac{\operatorname{i}}{2}\sqrt{\frac{7}{33}}\left[\operatorname{Im}(A_1A^*_0)-6\sqrt{\frac{5}{2}}\operatorname{Im}(A_2A^*_1)\right.\nonumber\\
        &\quad\left.+5\sqrt{3}\operatorname{Im}(A_3A^*_2)-5\sqrt{2}\operatorname{Im}(A_4A^*_3)\right]\\
   \beta^{22}_{1-}&=\frac{5\operatorname{i}}{2}\sqrt{\frac{5}{7}}\left[\operatorname{Im}(A_1A^*_0)+\frac{1}{\sqrt{10}}\operatorname{Im}(A_2A^*_1)\right.\nonumber\\
        &\quad\left.-\frac{1}{5\sqrt{3}}\operatorname{Im}(A_3A^*_2)+\frac{\sqrt{2}}{5}\operatorname{Im}(A_4A^*_3)\right]\\
   \beta^{42}_{1-}&=\frac{15\operatorname{i}}{\sqrt{77}}\left[\operatorname{Im}(A_1A^*_0)-\frac{4}{3\sqrt{10}}\operatorname{Im}(A_2A^*_1)\right.\nonumber\\
        &\quad\left.+\frac{1}{2\sqrt{3}}\operatorname{Im}(A_3A^*_2)-\frac{1}{\sqrt{2}}\operatorname{Im}(A_4A^*_3)\right]\\
   \beta^{62}_{1-}&=\frac{\operatorname{i}}{2}\sqrt{\frac{5}{11}}\left[\operatorname{Im}(A_1A^*_0)-\sqrt{\frac{5}{2}}\operatorname{Im}(A_2A^*_1)\right.\nonumber\\
        &\quad\left.-\frac{5}{\sqrt{3}}\operatorname{Im}(A_3A^*_2)+5\sqrt{2}\operatorname{Im}(A_4A^*_3)\right]\\
   \beta^{23}_{1-}&=-\frac{5\operatorname{i}}{2\sqrt{2}}\left[\operatorname{Im}(A_1A^*_0)-\sqrt{\frac{2}{5}}\operatorname{Im}(A_2A^*_1)\right.\nonumber\\
        &\quad\left.-\frac{2}{5\sqrt{3}}\operatorname{Im}(A_3A^*_2)-\frac{\sqrt{2}}{5}\operatorname{Im}(A_4A^*_3)\right]\\
   \beta^{43}_{1-}&=-\frac{3\operatorname{i}}{2}\sqrt{\frac{10}{11}}\left[\operatorname{Im}(A_1A^*_0)
        +\frac{4}{3}\sqrt{\frac{2}{5}}\operatorname{Im}(A_2A^*_1)\right.\nonumber\\
        &\quad\left.+\frac{1}{\sqrt{3}}\operatorname{Im}(A_3A^*_2)+\frac{1}{\sqrt{2}}\operatorname{Im}(A_4A^*_3)\right]\\
   \beta^{63}_{1-}&=-\frac{\operatorname{i}}{2}\sqrt{\frac{7}{22}}\left[\operatorname{Im}(A_1A^*_0)+\sqrt{10}\operatorname{Im}(A_2A^*_1)\right.\nonumber\\
        &\quad\left.-\frac{10}{\sqrt{3}}\operatorname{Im}(A_3A^*_2)-5\sqrt{2}\operatorname{Im}(A_4A^*_3)\right]\\
   \beta^{24}_{1-}&=\frac{5\operatorname{i}}{2}\sqrt{\frac{5}{42}}\left[\operatorname{Im}(A_1A^*_0)-3\sqrt{\frac{2}{5}}\operatorname{Im}(A_2A^*_1)\right.\nonumber\\
        &\quad\left.+\frac{2\sqrt{3}}{5}\operatorname{Im}(A_3A^*_2)+\frac{\sqrt{2}}{5}\operatorname{Im}(A_4A^*_3)\right]\\
   \beta^{44}_{1-}&=\frac{5\operatorname{i}}{2}\sqrt{\frac{6}{7}}\left[\operatorname{Im}(A_1A^*_0)+4\sqrt{\frac{2}{5}}\operatorname{Im}(A_2A^*_1)\right.\nonumber\\
        &\quad\left.-\sqrt{3}\operatorname{Im}(A_3A^*_2)-\frac{1}{\sqrt{2}}\operatorname{Im}(A_4A^*_3)\right]\\
   \beta^{64}_{1-}&=\frac{\operatorname{i}}{2}\sqrt{\frac{5}{66}}\left[\operatorname{Im}(A_1A^*_0)+3\sqrt{10}\operatorname{Im}(A_2A^*_1)\right.\nonumber\\
        &\quad\left.+10\sqrt{3}\operatorname{Im}(A_3A^*_2)+5\sqrt{2}\operatorname{Im}(A_4A^*_3)\right]\\
   \beta^{22}_{2+}&=\frac{5}{\sqrt{21}}\Bigg[\operatorname{Re}(A_2A^*_0)\nonumber\\
        &\quad+\sqrt{3}\operatorname{Re}(A_3A^*_1)+2\sqrt{\frac{3}{5}}\operatorname{Re}(A_4A^*_2)\Bigg]\\
   \beta^{42}_{2+}&=3\sqrt{\frac{30}{77}}\Bigg[\operatorname{Re}(A_2A^*_0)\nonumber\\
        &\quad-\frac{1}{2\sqrt{3}}\operatorname{Re}(A_3A^*_1)-\frac{2}{3}\sqrt{\frac{5}{3}}\operatorname{Re}(A_4A^*_2)\Bigg]\\
   \beta^{62}_{2+}&=\sqrt{\frac{10}{33}}\left[\operatorname{Re}(A_2A^*_0)\right.\nonumber\\
        &\quad\left.-2\sqrt{3}\operatorname{Re}(A_3A^*_1)+\sqrt{15}\operatorname{Re}(A_4A^*_2)\right]\\
   \beta^{23}_{2+}&=-\frac{5}{2\sqrt{3}}\left[\operatorname{Re}(A_2A^*_0)-2\sqrt{\frac{3}{5}}\operatorname{Re}(A_4A^*_2)\right]\\
   \beta^{43}_{2+}&=-\frac{3}{2}\sqrt{\frac{30}{11}}
    \left[\operatorname{Re}(A_2A^*_0)+\frac{2}{3}\sqrt{\frac{5}{3}}\operatorname{Re}(A_4A^*_2)\right]\\
   \beta^{63}_{2+}&=-\frac{1}{2}\sqrt{\frac{70}{33}}\left[\operatorname{Re}(A_2A^*_0)-\sqrt{15}\operatorname{Re}(A_4A^*_2)\right]\\
   \beta^{24}_{2+}&=\frac{5}{2\sqrt{7}}\Bigg[\operatorname{Re}(A_2A^*_0)\nonumber\\
        &\quad-\frac{4}{\sqrt{3}}\operatorname{Re}(A_3A^*_1)
        +2\sqrt{\frac{3}{5}}\operatorname{Re}(A_4A^*_2)\Bigg]\\
   \beta^{44}_{2+}&=\frac{9}{2}\sqrt{\frac{10}{77}}\Bigg[\operatorname{Re}(A_2A^*_0)\nonumber\\
        &\quad+\frac{2\sqrt{3}}{9}\operatorname{Re}(A_3A^*_1)-\frac{2\sqrt{15}}{9}\operatorname{Re}(A_4A^*_2)\Bigg]\\
   \beta^{64}_{2+}&=\frac{1}{2}\sqrt{\frac{10}{11}}\Bigg[\operatorname{Re}(A_2A^*_0)\nonumber\\
        &\quad+\frac{8}{\sqrt{3}}\operatorname{Re}(A_3A^*_1)+\sqrt{15}\operatorname{Re}(A_4A^*_2)\Bigg]\\
   \beta^{22}_{2-}&=\frac{5\operatorname{i}}{\sqrt{21}}\Bigg[\operatorname{Im}(A_2A^*_0)\nonumber\\
        &\quad+\sqrt{3}\operatorname{Im}(A_3A^*_1)+2\sqrt{\frac{3}{5}}\operatorname{Im}(A_4A^*_2)\Bigg]\\
   \beta^{42}_{2-}&=3\operatorname{i}\sqrt{\frac{30}{77}}\Bigg[\operatorname{Im}(A_2A^*_0)\nonumber\\
        &\quad-\frac{1}{2\sqrt{3}}\operatorname{Im}(A_3A^*_1)
        -\frac{2}{3}\sqrt{\frac{5}{3}}\operatorname{Im}(A_4A^*_2)\Bigg]\\
   \beta^{62}_{2-}&=\operatorname{i}\sqrt{\frac{10}{33}}\Bigg[\operatorname{Im}(A_2A^*_0)\nonumber\\
        &\quad-2\sqrt{3}\operatorname{Im}(A_3A^*_1)+\sqrt{15}\operatorname{Im}(A_4A^*_2)\Bigg]\\
   \beta^{23}_{2-}&=-\frac{5\operatorname{i}}{2\sqrt{3}}\left[\operatorname{Im}(A_2A^*_0)-2\sqrt{\frac{3}{5}}\operatorname{Im}(A_4A^*_2)\right]\\
   \beta^{43}_{2-}&=-\frac{3\operatorname{i}}{2}\sqrt{\frac{30}{11}}
    \left[\operatorname{Im}(A_2A^*_0)+\frac{2}{3}\sqrt{\frac{5}{3}}\operatorname{Im}(A_4A^*_2)\right]\\
   \beta^{63}_{2-}&=-\frac{\operatorname{i}}{2}\sqrt{\frac{70}{33}}\left[\operatorname{Im}(A_2A^*_0)-\sqrt{15}\operatorname{Im}(A_4A^*_2)\right]\\
   \beta^{24}_{2-}&=\frac{5\operatorname{i}}{2\sqrt{7}}\Bigg[\operatorname{Im}(A_2A^*_0)\nonumber\\
        &\quad-\frac{4}{\sqrt{3}}\operatorname{Im}(A_3A^*_1)
        +2\sqrt{\frac{3}{5}}\operatorname{Im}(A_4A^*_2)\Bigg]\\
   \beta^{44}_{2-}&=\frac{9\operatorname{i}}{2}\sqrt{\frac{10}{77}}\Bigg[\operatorname{Im}(A_2A^*_0)\nonumber\\
        &\quad+\frac{2\sqrt{3}}{9}\operatorname{Im}(A_3A^*_1)-\frac{2\sqrt{15}}{9}\operatorname{Im}(A_4A^*_2)\Bigg]\\
   \beta^{64}_{2-}&=\frac{\operatorname{i}}{2}\sqrt{\frac{10}{11}}\Bigg[\operatorname{Im}(A_2A^*_0)\nonumber\\
        &\quad+\frac{8}{\sqrt{3}}\operatorname{Im}(A_3A^*_1)+\sqrt{15}\operatorname{Im}(A_4A^*_2)\Bigg]\\
   \beta^{43}_{3+}&=-\frac{3}{2}\sqrt{\frac{35}{11}}
    \left[\operatorname{Re}(A_3A^*_0)+\frac{\sqrt{2}}{3}\operatorname{Re}(A_4A^*_1)\right]\\
   \beta^{63}_{3+}&=-\frac{1}{2}\sqrt{\frac{70}{11}}
    \left[\operatorname{Re}(A_3A^*_0)-\frac{3}{\sqrt{2}}\operatorname{Re}(A_4A^*_1)\right]\\
   \beta^{44}_{3+}&=\frac{3}{2}\sqrt{\frac{35}{11}}
    \left[\operatorname{Re}(A_3A^*_0)-\frac{\sqrt{2}}{3}\operatorname{Re}(A_4A^*_1)\right]\\
   \beta^{64}_{3+}&=\frac{1}{2}\sqrt{\frac{70}{11}}
    \left[\operatorname{Re}(A_3A^*_0)+\frac{3}{\sqrt{2}}\operatorname{Re}(A_4A^*_1)\right]\\
   \beta^{43}_{3-}&=-\frac{3\operatorname{i}}{2}\sqrt{\frac{35}{11}}
    \left[\operatorname{Im}(A_3A^*_0)+\frac{\sqrt{2}}{3}\operatorname{Im}(A_4A^*_1)\right]\\
   \beta^{63}_{3-}&=-\frac{\operatorname{i}}{2}\sqrt{\frac{70}{11}}
    \left[\operatorname{Im}(A_3A^*_0)-\frac{3}{\sqrt{2}}\operatorname{Im}(A_4A^*_1)\right]\\
   \beta^{44}_{3-}&=\frac{3\operatorname{i}}{2}\sqrt{\frac{35}{11}}
    \left[\operatorname{Im}(A_3A^*_0)-\frac{\sqrt{2}}{3}\operatorname{Im}(A_4A^*_1)\right]\\
   \beta^{64}_{3-}&=\frac{\operatorname{i}}{2}\sqrt{\frac{70}{11}}
    \left[\operatorname{Im}(A_3A^*_0)+\frac{3}{\sqrt{2}}\operatorname{Im}(A_4A^*_1)\right]\\
   \beta^{44}_{4+}&=\sqrt{\frac{105}{11}}\operatorname{Re}(A_4A^*_0)\\
   \beta^{64}_{4+}&=\frac{5}{2}\sqrt{\frac{14}{11}}\operatorname{Re}(A_4A^*_0)\\
   \beta^{44}_{4-}&=\operatorname{i}\sqrt{\frac{105}{11}}\operatorname{Im}(A_4A^*_0)\\
   \beta^{64}_{4-}&=\frac{5\operatorname{i}}{2}\sqrt{\frac{14}{11}}\operatorname{Im}(A_4A^*_0)
\end{align}

\subsection{Expressions of $\gamma^{J_3J_2}_{d'\pm}$}
\label{sec:apdix subsec4}
\begin{align}
   \gamma^{00}_{0+}&=1\\
   \gamma^{01}_{0+}&=\frac{1}{\sqrt{2}}(|E_0|^2+2|E_2|^2)\\
   \gamma^{02}_{0+}&=-\sqrt{\frac{10}{7}}\left(|E_0|^2+\frac{1}{2}|E_1|^2-|E_2|^2\right)\\
   \gamma^{03}_{0+}&=-\sqrt{2}\left(|E_1|^2+\frac{1}{2}|E_2|^2\right)\\
   \gamma^{04}_{0+}&=3\sqrt{\frac{2}{7}}\left(|E_0|^2-\frac{2}{3}|E_1|^2+\frac{1}{6}|E_2|^2\right)\\
   \gamma^{10}_{0+}&=-\sqrt{\frac{3}{2}}(|E_0|^2-|E_2|^2)\\
   \gamma^{11}_{0+}&=\sqrt{3}|E_2|^2\\
   \gamma^{12}_{0+}&=\sqrt{\frac{15}{7}}(|E_0|^2+|E_2|^2)\\
   \gamma^{13}_{0+}&=\frac{\sqrt{3}}{2}|E_2|^2\\
   \gamma^{14}_{0+}&=-3\sqrt{\frac{3}{7}}(|E_0|^2-\frac{1}{6}|E_2|^2)\\
   \gamma^{20}_{0+}&=\frac{1}{\sqrt{2}}(|E_0|^2-2|E_1|^2+|E_2|^2)\\
   \gamma^{21}_{0+}&=-(|E_1|^2-|E_2|^2)\\
   \gamma^{22}_{0+}&=-\sqrt{\frac{5}{7}}(|E_1|^2-|E_1|^2-|E_2|^2)\\
   \gamma^{23}_{0+}&=2\left(|E_1|^2+\frac{1}{4}|E_2|^2\right)\\
   \gamma^{24}_{0+}&=\frac{3}{\sqrt{7}}\left(|E_0|^2+\frac{4}{3}|E_1|^2+\frac{1}{6}|E_2|^2\right)\\
   \gamma^{11}_{1+}&=\frac{3}{2}\left[\operatorname{Re}(E_1E^*_0)+\sqrt{\frac{2}{3}}\operatorname{Re}(E_2E^*_1)\right]\\
   \gamma^{12}_{1+}&=\frac{1}{2}\sqrt{\frac{15}{7}}\left[\operatorname{Re}(E_1E^*_0)+\sqrt{6}\operatorname{Re}(E_2E^*_1)\right]\\
   \gamma^{13}_{1+}&=-\sqrt{\frac{3}{2}}\left[\operatorname{Re}(E_1E^*_0)-\sqrt{\frac{3}{2}}\operatorname{Re}(E_2E^*_1)\right]\\
   \gamma^{14}_{1+}&=-\frac{3}{2}\sqrt{\frac{10}{7}}
    \left[\operatorname{Re}(E_1E^*_0)-\frac{1}{\sqrt{6}}\operatorname{Re}(E_2E^*_1)\right]\\
   \gamma^{21}_{1+}&=-\frac{3}{2}\left[\operatorname{Re}(E_1E^*_0)-\sqrt{\frac{2}{3}}\operatorname{Re}(E_2E^*_1)\right]\\
   \gamma^{22}_{1+}&=-\frac{1}{2}\sqrt{\frac{15}{7}}\left[\operatorname{Re}(E_1E^*_0)-\sqrt{6}\operatorname{Re}(E_2E^*_1)\right]\\
   \gamma^{23}_{1+}&=\sqrt{\frac{3}{2}}\left[\operatorname{Re}(E_1E^*_0)+\sqrt{\frac{3}{2}}\operatorname{Re}(E_2E^*_1)\right]\\
   \gamma^{24}_{1+}&=\frac{3}{2}\sqrt{\frac{10}{7}}
    \left[\operatorname{Re}(E_1E^*_0)+\frac{1}{\sqrt{6}}\operatorname{Re}(E_2E^*_1)\right]\\
   \gamma^{11}_{1-}&=\frac{3\operatorname{i}}{2}\left[\operatorname{Im}(E_1E^*_0)+\sqrt{\frac{2}{3}}\operatorname{Im}(E_2E^*_1)\right]\\
   \gamma^{12}_{1-}&=\frac{\operatorname{i}}{2}\sqrt{\frac{15}{7}}\left[\operatorname{Im}(E_1E^*_0)+\sqrt{6}\operatorname{Im}(E_2E^*_1)\right]\\
   \gamma^{13}_{1-}&=-\operatorname{i}\sqrt{\frac{3}{2}}\left[\operatorname{Im}(E_1E^*_0)-\sqrt{\frac{3}{2}}\operatorname{Im}(E_2E^*_1)\right]\\
   \gamma^{14}_{1-}&=-\frac{3\operatorname{i}}{2}\sqrt{\frac{10}{7}}
    \left[\operatorname{Im}(E_1E^*_0)-\frac{1}{\sqrt{6}}\operatorname{Im}(E_2E^*_1)\right]\\
   \gamma^{21}_{1-}&=-\frac{3\operatorname{i}}{2}\left[\operatorname{Im}(E_1E^*_0)-\sqrt{\frac{2}{3}}\operatorname{Im}(E_2E^*_1)\right]\\
   \gamma^{22}_{1-}&=-\frac{\operatorname{i}}{2}\sqrt{\frac{15}{7}}\left[\operatorname{Im}(E_1E^*_0)-\sqrt{6}\operatorname{Im}(E_2E^*_1)\right]\\
   \gamma^{23}_{1-}&=\operatorname{i}\sqrt{\frac{3}{2}}\left[\operatorname{Im}(E_1E^*_0)+\sqrt{\frac{3}{2}}\operatorname{Im}(E_2E^*_1)\right]\\
   \gamma^{24}_{1-}&=\frac{3\operatorname{i}}{2}\sqrt{\frac{10}{7}}
    \left[\operatorname{Im}(E_1E^*_0)+\frac{1}{\sqrt{6}}\operatorname{Im}(E_2E^*_1)\right]\\
   \gamma^{22}_{2+}&=\sqrt{\frac{30}{7}}\operatorname{Re}(E_2E^*_0)\\
   \gamma^{23}_{2+}&=\sqrt{\frac{15}{2}}\operatorname{Re}(E_2E^*_0)\\
   \gamma^{24}_{2+}&=\frac{3}{2}\sqrt{\frac{10}{7}}\operatorname{Re}(E_2E^*_0)\\
   \gamma^{22}_{2-}&=\operatorname{i}\sqrt{\frac{30}{7}}\operatorname{Im}(E_2E^*_0)\\
   \gamma^{23}_{2-}&=\operatorname{i}\sqrt{\frac{15}{2}}\operatorname{Im}(E_2E^*_0)\\
   \gamma^{24}_{2-}&=\frac{3\operatorname{i}}{2}\sqrt{\frac{10}{7}}\operatorname{Im}(E_2E^*_0)
\end{align}


\begin{thebibliography}{99}


\bibitem{Higher charmonia}
    T. Barnes, S. Godfrey, E.S. Swanson,
    Phys. Rev. D
    \href{http://arxiv.org/abs/hep-ph/0505002}{{\bf 72}, 054026 (2005)}

\bibitem{New states above charm threshold}
    E. Eichten, K. Lane, C. Quigg,
    Phys. Rev. D
    \href{http://link.aps.org/doi/10.1103/PhysRevD.73.014014}{{\bf 73}, 014014 (2006)} 

\bibitem{Quarkonia and their transitions}
    E. Eichten, S. Godfrey, H. Mahlke, J. Rosner,
    Rev. Mod. Phys.
    \href{http://link.aps.org/doi/10.1103/RevModPhys.80.1161}{{\bf 80}, 1161 (2008)}

\bibitem{Charmonium spectroscopy above thresholds}
    T. Fern\'{a}ndez-Caram\'{e}s, A. Valcarce, J. Vijande,
    Phys. Rev. Lett.
    \href{http://link.aps.org/doi/10.1103/PhysRevLett.103.222001}{{\bf 103}, 222001 (2009)}

\bibitem{Charmonium options for the X(3872)}
    T. Barnes, S. Godfrey,
    Phys. Rev. D
    \href{http://link.aps.org/doi/10.1103/PhysRevD.69.054008}{{\bf 69}, 054008 (2004)}

\bibitem{Charmonium levels near threshold and the narrow state}
    E. Eichten, K. Lane, C. Quigg,
    Phys. Rev. D
    \href{http://link.aps.org/doi/10.1103/PhysRevD.69.094019}{{\bf 69}, 094019 (2004)}

\bibitem{The PANDA experiment at FAIR}
    D. Bettoni,
    {\it Proc. CHARM 2007 Workshop} (New York, 2007), \href{http://arxiv.org/abs/0710.5664}{hep-ex/0710.5664v1}

\bibitem{Physics performance report for PANDA: strong interaction studies with antiprotons}
    \={P}ANDA Collaboration,
    {\it Physics performance report for PANDA: strong interaction studies with antiprotons}, 2009, \href{http://arxiv.org/abs/0903.3905}{hep-ex/0903.3905v1}

\bibitem{previous paper}
    A.W.K. Mok, C.P. Wong, W.Y. Sit,
    J. High Energy Phys.
    \href{http://dx.doi.org/10.1007/JHEP10%282012%29083}{{\bf 1210}, 083 (2012)}


\bibitem{Mok 2010}
    A.W.K. Mok, K.J. Sebastian,
    Eur. Phys. J. C
    \href{http://dx.doi.org/10.1140/epjc/s10052-010-1289-3}{{\bf 67}, 125 (2010)}

\bibitem{Mok 2011}
    A.W.K. Mok, M.F. Chow,
    Eur. Phys. J. C
    \href{http://dx.doi.org/10.1140/epjc/s10052-011-1792-1}{{\bf 71}, 1792 (2011)}

\bibitem{Mok 2008}
    A.W.K. Mok, K.J. Sebastian,
    Eur. Phys. J. C
    \href{http://dx.doi.org/10.1140/epjc/s10052-008-0649-8}{{\bf 56}, 189 (2008)}

\bibitem{Martin}
    A.D. Martin, T.D. Spearman,
    {\it Elementary Particle Theory} (North-Holland, Amsterdam, 1970), p. 113-115

\bibitem{KJ Sebastian}
    K.J. Sebastian,
    Phys. Rev. D
    \href{http://link.aps.org/doi/10.1103/PhysRevD.49.3450}{{\bf 49}, 3450 (1994)}

\bibitem{Mok 2009}
    A.W.K. Mok, K.J. Sebastian,
    Eur. Phys. J. C
    \href{http://dx.doi.org/10.1140/epjc/s10052-009-1091-2}{{\bf 63}, 101 (2009)}

\bibitem{phys rev D13 1203 (1976)}
    F. Karl, S. Meshkov, J.L. Rosner,
    Phys. Rev. D
    \href{http://link.aps.org/doi/10.1103/PhysRevD.13.1203}{{\bf 13}, 1203 (1976)}

\bibitem{phys rev D55 225 (1997)}
    K.J. Sebastian, X.G. Zhang,
    Phys. Rev. D
    \href{http://link.aps.org/doi/10.1103/PhysRevD.55.225}{{\bf 55}, 225 (1997)}

\end{thebibliography}
\end{document}